\documentclass[sigconf]{acmart}

\usepackage{csquotes}
\usepackage{enumitem}

\usepackage{subcaption}
\usepackage{mwe}
\usepackage{wrapfig}
\usepackage[flushleft]{threeparttable}
\usepackage{multirow}

\usepackage{epstopdf}
\usepackage{booktabs}
\usepackage{comment}
\usepackage[np, autolanguage]{numprint}

\AtBeginDocument{%
  \providecommand\BibTeX{{%
    \normalfont B\kern-0.5em{\scshape i\kern-0.25em b}\kern-0.8em\TeX}}}

\setcopyright{acmcopyright}
\copyrightyear{2018}
\acmYear{2018}
\acmDOI{10.1145/1122445.1122456}

\acmConference[FAccT '21]{FAccT '21: ACM Conference on Fairness, Accountability, and Transparency}{March, 2021}{}
\acmBooktitle{FAccT '21: ACM Conference on Fairness, Accountability, and Transparency, March 2021}
\acmPrice{15.00}
\acmISBN{978-1-4503-XXXX-X/18/06}

\begin{document}


\title{Operationalizing Framing to Support Multiperspective Recommendations of Opinion Pieces}

\settopmatter{authorsperrow=4}

\author{Mats Mulder}
\authornote{Work done while enrolled in a master program at Delft University of Technology.}
\affiliation{%
  \institution{Delft University of Technology}
  \city{Delft}
  \country{The Netherlands}
}
\email{mats.mulder@live.nl}

\author{Oana Inel}
\affiliation{%
  \institution{Delft University of Technology}
  \city{Delft}
  \country{The Netherlands}}
\email{o.inel@tudelft.nl}

\author{Jasper Oosterman}
\affiliation{%
  \institution{Blendle}
  \city{Utrecht}
  \country{The Netherlands}
}
\email{jasperoosterman@blendle.com}

\author{Nava Tintarev}
\affiliation{%
 \institution{Maastricht University }
 \city{Maastricht\\}
 \country{The Netherlands}
}
\email{n.tintarev@maastrichtuniversity.nl}

\renewcommand{\shortauthors}{Mulder et al.}

\newcommand{\Problem}{\colorbox{blue!30}{~Problem Definition~}}
\newcommand{\Attribution}{\colorbox{gray!30}{~Causal Attribution~}}
\newcommand{\Evaluation}{\colorbox{purple!30}{~Moral Evaluation~}}
\newcommand{\Recommendation}{\colorbox{red!30}{~Treatment Recommendation~}}

\begin{abstract}

Diversity in personalized news recommender systems is often defined as dissimilarity, and based on topic diversity (\emph{e.g.}, corona versus farmers strike). Diversity in news media, however, is understood as multiperspectivity (\emph{e.g.}, different opinions on corona measures), and arguably a key responsibility of the press in a democratic society.
While viewpoint diversity is often considered synonymous with source diversity in communication science domain, in this paper, we take a computational view. We operationalize the notion of framing, adopted from communication science. We apply this notion to a re-ranking of topic-relevant recommended lists, to form the basis of a novel viewpoint diversification method. Our offline evaluation indicates that the proposed method is capable of enhancing the viewpoint diversity of recommendation lists according to a diversity metric from literature. In an online study, on the Blendle platform, a Dutch news aggregator platform, with more than 2000 users, we found that \textit{users are willing to consume viewpoint diverse news recommendations}. We also found that \textit{presentation characteristics} significantly influence the reading behaviour of diverse recommendations. These results suggest that future research on presentation aspects of recommendations can be just as important as novel viewpoint diversification methods to truly achieve multiperspectivity in online news environments.
\end{abstract}

\ccsdesc[500]{Information systems~Recommender systems}
\ccsdesc[500]{Human-centered computing~User studies}
\ccsdesc[500]{Human-centered computing~Empirical studies in HCI}

\keywords{recommender systems, viewpoint diversity, framing aspects}

\newcommand{\nt}[1]{\textcolor{blue}{\textbf{/* #1 (Nava) */}}}

\newcommand{\oi}[1]{\textcolor{purple}{\textbf{/* #1 (Oana) */}}}

\maketitle


\section{Introduction}
\label{sec:introduction}

In recent years, traditional news sources are increasingly using online news platforms to distribute their content. Digital-born news websites and \textit{news aggregators}, which combine content from various sources in one service, are also gaining ground \cite{newman2015reuters}. In 2015, 23\% of survey respondents reported online media as their primary news source, and 44\% considered digital and traditional sources equally relevant \cite{newman2015reuters}. This change also induces a wide adoption of \textit{news recommender systems} that automatically provide personalized news recommendations to users.

Communication studies generally acknowledge two important roles of media in a democratic society \cite{helberger2019democratic}. The first role is to inform citizens about important societal and political issues. The second role is to foster a diverse public sphere. Both roles are then related to multiple social-cultural objectives of democracy, such as informed decision-making, cultural pluralism and citizens welfare \cite{mcquail1992media,stromback2005search}.

The role of news recommender systems in promoting these democratic values is under heavy discussion in academic debate. For example, the term \textit{filter bubbles} received increasing awareness, suggesting that high levels of personalisation would lock people up people in bubbles of what they already know or think \cite{pariser2011filter}. According to \citeauthor{helberger2019democratic}, the democratic role of news recommender systems mainly depends on the democratic theory that is being followed. In their conceptual framework, this role is being evaluated for the most common theories: the liberal, the participatory and the deliberative \cite{helberger2019democratic}. In particular in relation to the participatory and deliberative model, the development of viewpoint diversification methods can be motivated.

However, current diversification methods \cite{kunaver2017diversity,ziegler2005improving} do not address viewpoint diversity, but define diversity as dissimilarity and operationalize it through topic diversity (\emph{e.g.}, corona versus farmers strike). Therefore, current diversification methods are not applicable in the news domain, and novel viewpoint diversification methods are needed to maintain and assure multiperspectivity in online news environments. To truly enable \emph{multiperspectivity}, users should be willing to consume viewpoint-diverse recommendations. Moreover, their behaviour should be studied in real, online scenarios. Thus, we investigate the following research questions:



\textbf{R1}: \textit{How is reading behaviour affected by viewpoint diverse news recommendations?}

\textbf{R2}: \textit{How is reading behaviour affected by presentation characteristics of viewpoint diverse news recommendations?}

To answer these questions, we propose a re-ranking approach for lists of recommended articles based on aspects of news frames, a concept taken from communication studies. In particular, a news frame describes how to identify a view on an issue, in a given article \cite{entman1993framing}. Thus, by bridging aspects from the social and the computational domains, we aim to overcome the current gap between the definition of diversity in recommender systems and news media. 

During an offline evaluation, the proposed method increased the viewpoint diversity of recommended lists of news articles on several topics. Further, we measured the influence of the viewpoint diversification method on the reading behaviour of more than 2000 users, which are likely to interact with the recommended articles, in an online study on the Blendle platform, a Dutch news aggregator platform. We found that reading behaviour of users that received diverse recommendations was comparable with the reading behaviour of users that received news articles optimized only for relevance. However, we did find 
a positive influence of two \textit{presentation characteristics} on the click-through rate of recommendations, \emph{i.e.}, news articles with thumbnails and news articles with more hearts are more often read. 
Therefore, we make the following contributions:

\begin{itemize}[noitemsep,topsep=0pt]
    \item a novel method for \textit{viewpoint diversification} using re-ranking of news recommendation lists, based on framing aspects; 
    \item an online evaluation with more than 2000 users, on the Blendle platform, to understand: 
    \begin{enumerate}[label=(\alph*),noitemsep,topsep=0pt]
        \item how viewpoint-diverse recommendations affect the reading behaviour of users; and
        \item how article's \textit{presentation characteristics} affect the reading behaviour of users.
    \end{enumerate}
\end{itemize}

\section{Related Work}
\label{sec:related_work}

In this section, we first investigate how communication science understands diversity. Then, we review current approaches for diversity in recommender systems. These allow us to bridge the gap between the domains of communication and computer science, by operationalizing framing aspects in a diversification algorithm. 

\subsection{Diversity in News Media}
In news media, diversity refers to multiperspectivity or a diversity of viewpoints \cite{gans2003democracy}. In communication science, diversity is, in general, a key measure for news quality~\cite{porto2007frame,choi2009diversity,masini2018measuring}, thus fostering multiple democratic aspects, such as informed decision-making, cultural pluralism and citizens welfare~\cite{napoli1999deconstructing,voakes1996diversity}. Two main approaches for assessing diversity can be distinguished: source and content diversity \cite{napoli1999deconstructing, baden2017conceptualizing, benson2009makes}, with most studies focusing on source diversity \cite{napoli1999deconstructing, baden2017conceptualizing, voakes1996diversity,baden2014ple}. When measuring source diversity, most methods follow \citet{bennett1996introduction}'s indexing theory, which assumes that including non-official or non-elite sources corresponds to high levels of diversity \cite{baden2017conceptualizing}. Alternatively, \citet{napoli1999deconstructing} approaches the issue from a policymaker point of view and distinguishes three aspects of source diversity: content ownership or programming, ownership of media outlets, and the workforce within individual media outlets.

Critics, however, state that multiple sources can still foster the same point of view and therefore, source diversity is not a direct measure for viewpoint diversity \cite{voakes1996diversity}. Multiple studies also indicate that power distributions in society, commercial pressure of news media and journalistic norms and practices, significantly influence which sources gain media access \cite{benson2009makes,baden2017conceptualizing}. Therefore, it is often argued that viewpoint diversity can only be achieved by fostering content diversity \cite{masini2018measuring,napoli1999deconstructing, choi2009diversity, gans2003democracy, baker2001media, voakes1996diversity}. Content diversity is defined in \cite{van1999competition} as \emph{``heterogeneity of media content in terms of one or more specified characteristics''}. \citet{baden2017conceptualizing} identified six common approaches to assess content diversity. The first three methods focus on the tone or political position represented in the news, \emph{i.e.}, the inclusion of non-official positions, the diversity of political tone or analysis of political slant. These methods, however, assume that political disagreement equals viewpoint diversity \cite{baden2017conceptualizing}. 
Another approach uses language diversity to evaluate content diversity. However, this is again no direct measure, since different language can describe the same perspective \cite{baden2017conceptualizing}.

The final two approaches use the concept of \textit{frames} to assess content diversity. Framing theory states that every communicative message selectively emphasizes certain aspects of the complex reality \cite{baden2017conceptualizing}. Thereby, frames enable different interpretations of the same issue \cite{scheufele1999framing}. Framing has been put forward by many scholars to enhance content diversity. For example, \citet{porto2007frame} states that news environments need to be evaluated by their ability to provide diverse frames. \citet{baden2017conceptualizing} describe three frames' aspects that are central to the role of viewpoint diversity in democratic media. First, frames create different interpretations of the same issue by selecting some aspects of the complex reality \cite{gamson1989media}. Second, frames are not neutral but suggest specific evaluations and courses of actions that serve some purpose better than other \cite{entman1993framing}. Third, frames are often strategically constructed to advocate particular political views and agendas. Framing, thus, can be a suitable conceptualization of viewpoint diversity.

\subsection{Diversity in Recommender Systems}
Traditionally, research on recommender systems focused on evaluating their performance in terms of accuracy metrics \cite{ziegler2005improving}. Such focus, however, induced a problem which is known as \textit{over-fitting}, \emph{e.g.}, a model is fitted so strongly to a user that it is unable to detect any other interests \cite{kunaver2017diversity}. Additionally, there is a need for a more user-centric evaluation of recommender systems. Thus, diversity has become one of the most prominent \textit{beyond-accuracy metrics} for recommender systems \cite{ziegler2005improving}. In this context, diversity is generally defined as the opposite of similarity \cite{kunaver2017diversity}, and it is often based on topic diversity (\emph{e.g.}, corona versus farmers strike). For example, \citet{ziegler2005improving} proposed a topic diversification method based in the \textit{intra-list diversity} metric.

Current diversification methods for recommender systems, thus, do not focus on viewpoint diversity and are not applicable in the news domain. To the best of our knowledge, only one study for viewpoint diversification has been proposed so far \cite{tintarev2018same}. \citet{tintarev2018same} propose a new distance measure for viewpoint diversity based on linguistic representations of news articles. This diversity measure was then applied in a post-processing re-ranking algorithm \cite{carbonell1998use} to a list of news articles. These allowed optimizing for the balance between topic relevance and viewpoint diversity. In a small scale user study \cite{tintarev2018same}, readers indicated a lower intent to consume diversified content, motivating the need to study behavioural measures for newsreaders on a larger scale. Thus, we argue that more research is required to understand the relationship between the metric and the influence on readers behaviour. 


In this work, we aim to bridge the current gap between the notion of framing in communication science and potential computational measure. Additionally, we aim to study how viewpoint diversification affects the behaviour of newsreaders in an applied setting. The next section justifies the operationalization of \emph{framing} in the computational domain.



\section{Framing for viewpoint diversity}
\label{sec:focus_group}

Framing is an extensively researched concept in different domains, including psychology, communication and sociology, having its roots in the latter domain.  \citet{bateson1955theory} state that communication only gets meaning in its context and by the way the message is constructed. Later, frame theory gained increasing momentum and was generally understood as follows: every communicative message selectively emphasizes certain aspects of a complex reality \cite{baden2017conceptualizing}. Thus, every news article (unintentionally) comprises some form of framing \cite{baden2017conceptualizing}. Frames are often deliberately used to construct strategic, often political, views on a topic. Consequently, frames enable different interpretations of the same issue \cite{baden2017conceptualizing}. However, every frame inevitably deselects other, equally plausible and relevant frame \cite{baden2017conceptualizing}.

When considering frames in news articles, multiple definitions exist \cite{giltin1980whole, gamson1989media, de2005news}. However, the definition of \citet{entman1993framing} is the most commonly adopted in the literature. It states that framing includes the selection of \emph{``some aspects of perceived reality and make the more salient in a communicating text, in such a way as to promote a particular definition of a problem, causal interpretation, moral evaluation and treatment recommendation for the item described''}. Within this definition, the problem describes \emph{four framing functions} - for which we also provide a running example -, namely:
\begin{enumerate}[noitemsep,topsep=0pt]
    \item \textbf{\Problem}: ``what a causal agent is doing with what costs and benefits''; \emph{e.g., a second Coronavirus wave is approaching};
    \item \textbf{\Attribution}: ``identifying the forces creating the problem''; \emph{e.g., (it is due to the) government policy response};
    \item \textbf{\Evaluation}: \emph{``evaluate causal agents and their effects''}; \emph{e.g.}, response to approaching second wave came too late (negative evaluation);
    \item \textbf{\Recommendation}: ``offer and justify treatments for the problems and predict their likely effects''; \emph{e.g., there must be predefined measures to be deployed at a critical threshold of virus spread.}
\end{enumerate} 

Additionally, \citet{entman1993framing} describes how to find frames at different levels of analysis, including single sentences, paragraphs or articles as a whole. Also, a frame may not necessarily include all the four functions. 

Most framing analysis approaches focus on manual analysis of articles \cite{kroon2016victims,matthes2008content,vliegenthart2012framing}. Only recently, some computer-assisted methods gained interest \cite{burscher2014teaching,vu2020news,greussing2017shifting}. As a result, the identification of frames often falls into a methodological black box \cite{matthes2008content}. Thereby, the main issue includes the ambiguity of \emph{``which elements should be present in an article or news story to signify the existence of a frame''} \cite{matthes2008content}. To overcome this problem, some recent studies \cite{matthes2008content,vliegenthart2012framing,baden2017conceptualizing} propose a novel identification method based on the extraction of the four aforementioned framing aspects in the definition of \citet{entman1993framing}. 

\subsection{Focus Group Setup}
To guide the operationalization of the framing aspects, we started with a qualitative analysis. Through a small focus group, we aimed to gain insights into how the four framing functions of the main frame of an article manifest in its content and how we can identify them computationally.
    
\subsubsection{Participants}
We invited three experts in the field of news article and framing analysis. All experts had a background in journalism, communication, or news media. They all had multiple years of relevant work experience. 

\subsubsection{Materials} As a basis for discussion during the focus group, we used opinion pieces on the topic of \emph{Dutch farmers protests}. Opinion pieces refer to news articles that reflect the authors opinion and thus, do not claim to be objective. An initial discussion with domain experts indicated that this type of news article is the most suitable to identify framing functions.

\subsubsection{Procedure}
The focus group procedure consisted of two steps. 

\emph{1. Annotation session:} First, the participants were asked to perform framing analysis on an opinion piece, using the four framing functions as described by \citet{entman1993framing}. In particular, the participants had to individually highlight parts of the article, such as word clauses or sentences, that can be related to one of the four framing functions of the main frame of the news article.

\emph{2. Review session:} Second, the results were discussed, together with some general questions on news article analysis and framing. For every highlighted part, we asked the participants to motivate why the highlighted part is related to one of the four framing functions. Besides, we used the results as input to a broader discussion on news article analysis and framing, such as:
    
\begin{itemize}[noitemsep,topsep=0pt]
    \item What is the main heuristic that you used to analyze the article?
    \item What procedure did you follow to analyze the framing functions of the article?
    \item Can you derive any patterns in the way framing functions manifest in opinion pieces?
\end{itemize}

    
\subsection{Results of Framing Analysis}
During the review session, all experts indicated that they used the article structure as the main heuristic to find the framing functions regarding the main frame. They also pointed out that opinion pieces are still strongly shaped by journalistic values on how an article should be structured. We further analyzed this heuristic according to the four framing functions: 

    \begin{enumerate}[noitemsep,topsep=0pt]
        \item \textbf{\Problem}:
        In opinion pieces, the first part of the article often presents the main problem that the author addresses and includes the title, the lead, and the first \textit{x} paragraphs. Work on manual frame analysis \cite{kroon2016victims} supports this finding. The number of introductory paragraphs, \textit{x}, can be different per source, author, or article. 
        \item \textbf{\Attribution + \Evaluation}:
        The body of an article is used to analyze the main problem and usually contains different factors that contribute to the problem under investigation and their evaluation. We can match this with: 
        a) the causal attribution of a frame (forces creating the problem), and b) the moral judgements (evaluate the causal attribution and their effect) \cite{entman1993framing}.
        \item \textbf{\Recommendation}:
        Treatment recommendations can be seen as suggestions to improve or solve the issue described by the problem definition of the main frame. They normally appear in the concluding paragraphs, according to the focus group members.
    \end{enumerate}
    
Note, however, that this structure is only a heuristic and it only applies to opinion pieces. Other types, such as interviews, are structured differently.

The results of the annotation session also indicate that each framing function related to the main frame of an article can normally be found within one paragraph. 
Additionally, a paragraph can include multiple framing functions, but words, clauses, and sentences generally represent a single framing function. 

\section{Dataset}
\label{sec:dataset}

In this section, we describe the experimental dataset, which consists of opinion pieces, in Dutch. The choice of article type is motivated by the focus group session presented in Section \ref{sec:focus_group}, in which the structure of this article type is put forward as the primary heuristic to find framing aspects. We picked topics that we expected \textbf{a)} to be present on the Blendle platform at the time when we performed the online user study; \textbf{b)} to contain different viewpoints addressed in the news; and \textbf{c)} to balance issues that more current versus long-standing. The dataset consists of four ongoing topics: \emph{Black Lives Matter}, \emph{Coronavirus}, \emph{U.S. Elections} - as more current topics, and the dominance and privacy issues around \emph{Big Tech} - as a long-standing topic. 


\begin{table*}[!ht]
\centering
\caption{Queries used (in Dutch) to retrieve news articles for the four topics in our dataset.}
\label{tab:data-selection-properties}
\resizebox{1\textwidth}{!}{
\begin{tabular}{lll}
\multicolumn{1}{c}{Topic}                                    & \multicolumn{1}{c}{Search Query} & \multicolumn{1}{c}{\begin{tabular}[c]{@{}c@{}}Start \\ Date\end{tabular}} \\ \toprule
\begin{tabular}[c]{@{}l@{}}Black Lives\\ Matter\end{tabular} & \begin{tabular}[c]{@{}l@{}}(’black lives matter’ OR ’racisme debat’ OR ’blm-demonstraties’ OR ’George Floyd’ OR ’racisme-debat’) AND NOT (’belastingdienst’ OR ’corona’)\end{tabular} & \begin{tabular}[c]{@{}l@{}}June 15\\ 2020\end{tabular} \\ \hline
Coronavirus & \begin{tabular}[c]{@{}l@{}}’corona’ OR ’covid-19’ OR ’mondkapjes’ OR ’mondkapje’ OR ’mondmasker’ OR ’mondkapjesplicht’ OR ’coronatest’ OR ’coronatesters’ OR ’rivm’ \\OR ’virus’ OR ’viroloog’ OR ’golf’ OR ’topviroloog’ OR ’uitbraak’ OR ’uitbraken’ OR ’coronaregels’ OR ’versoeplingen’ OR ’staatssteun’ OR ’vaccin’\end{tabular} & \begin{tabular}[c]{@{}l@{}}June 1\\ 2020\end{tabular} \\ \hline
U.S. Elections & \begin{tabular}[c]{@{}l@{}}’Donald Trump’ AND (’presidentsverkiezingen’ OR ’Verkiezingen’ OR ’campagne’ OR ’verkiezingsstrijd’ OR ’verkiezingscampagne’ OR ’Joe biden’)\end{tabular} & \begin{tabular}[c]{@{}l@{}}June 1\\ 2020\end{tabular} \\ \hline
Big Tech & \begin{tabular}[c]{@{}l@{}}(’macht’ OR ’machtig’ OR 'privacy’ OR ’data’ OR ’privacyonderzoek' OR privacy-schandaal’) AND (’big tech’ OR ’tech-bedrijven’  OR techbedrijven’)\end{tabular} & 2018  \\ \bottomrule                                                            
\end{tabular}
}
\end{table*}

We collected our dataset from an archive containing more than 5 million Dutch news articles. The archive is known to undergo checks for articles quality, to remove undesirable content, such as the weather or short actualities. For each topic, we used the search terms (queries) and restrictions shown in Table \ref{tab:data-selection-properties}. We provide the list of search terms in their original language, Dutch, because we do not want to add additional bias through translation. Additionally, since the proposed method heavily relies on the structure of the article, we set up a filter for the minimum number of words to 450 and a filter for the minimum number of paragraphs to 5. 



Table \ref{table:data-offline-properies} provides an overview of the dataset, per topic. 
While the length of the articles varies across topics, they are usually far longer than the 450-word limit we chose. Four publishers are present for all topics: De Volkskrant, De Standaard, Trouw and Het Algemeen Dagblad. Furthermore, De Volkskrant is the most prominent publisher for all topics, except for the \emph{U.S. Elections} topic. The inclusion of other, less frequent, publishers varies per topic. Overall, our dataset covers a set of 15 unique publishers.

We also present some properties concerning the presentation characteristics of the articles on the news aggregator website. We observe that the ratio of articles that contains a thumbnail image depends on the topic. For the \emph{Black Lives Matter} and \emph{Coronavirus} topics, more than half of the articles have a thumbnail image, while the opposite holds for the other two topics. The number of custom titles from the editorial team and the average title length also differ considerable per topic. Only a few articles have an editorial title, and they usually appear for the \emph{Big Tech} and \emph{U.S. Elections} topics.

\begin{table}[!ht]
\centering
\caption{Overview of the experimental dataset, per topic.}
\label{table:data-offline-properies}
\resizebox{1\columnwidth}{!}{
\begin{tabular}{lcccccc}
\multirow{2}{*}{Topic} & \multirow{2}{*}{Articles} & \multirow{2}{*}{Publishers} & \begin{tabular}[c]{@{}c@{}}Avg \\\#Words\end{tabular} & \begin{tabular}[c]{@{}c@{}}With \\ thumb.\end{tabular} & \begin{tabular}[c]{@{}c@{}}With \\ ed. title\end{tabular} & \begin{tabular}[c]{@{}c@{}}Avg title \\ length\end{tabular} \\ \toprule
\begin{tabular}[c]{@{}l@{}}Black Lives \\ Matter\end{tabular} & 69 & 10 & 697  & 39 & 1 & 6.3 \\
Coronavirus & 52 & 7 & 608 & 27 & 4 & 5.2 \\
U.S. Elections & 42 & 6 & 744 & 20 & 8 & 9.6 \\
Big Tech & 51 & 10 & 761 & 17 & 10 & 8.1 \\ \bottomrule
\end{tabular}
}
\end{table}

\section{Viewpoint Diversity Methodology}
\label{sec:methodology}

\begin{figure}[ht!]
    \centering
    \includegraphics[width=0.5\textwidth]{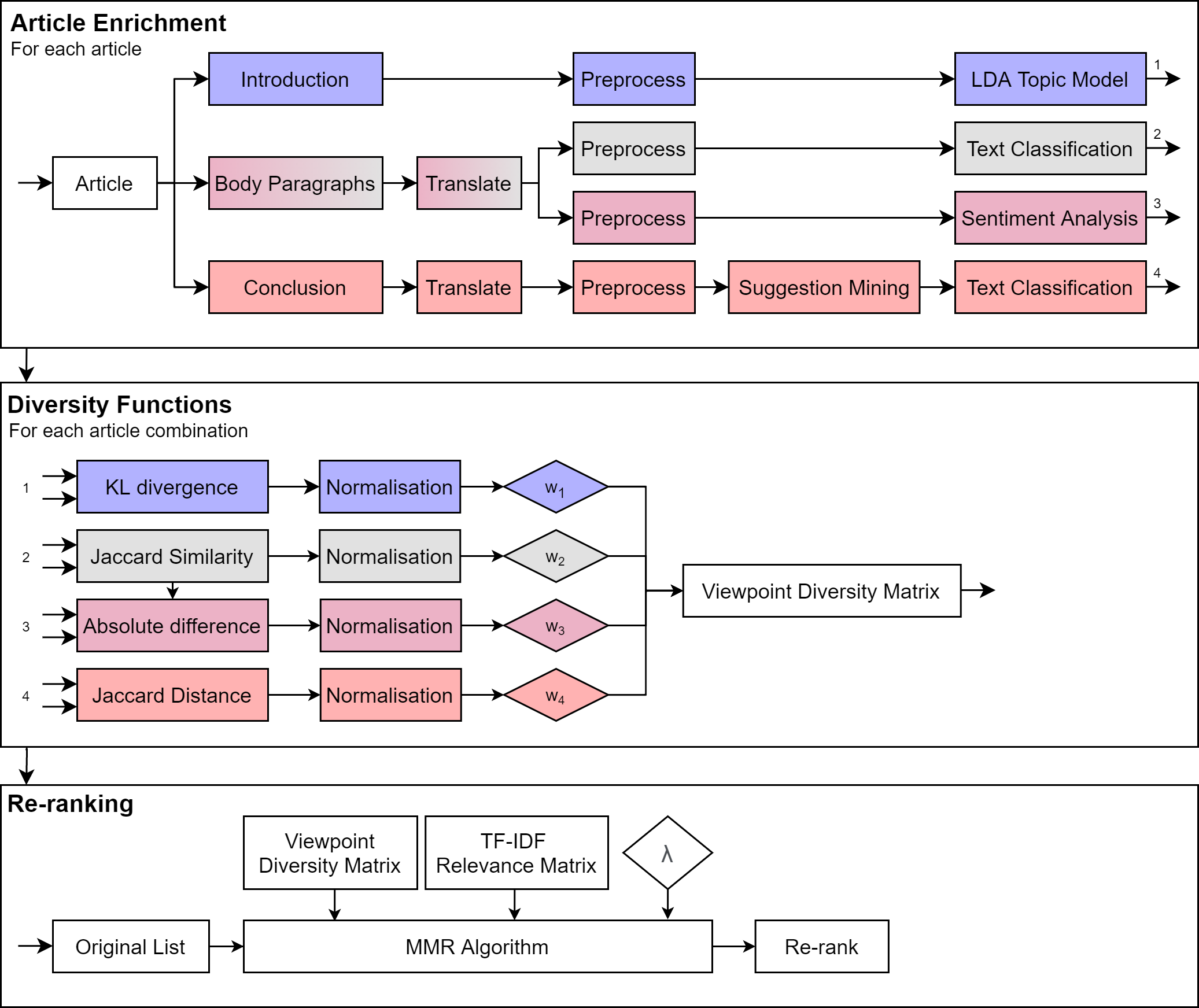}
    \caption{Viewpoint diversification pipeline}
    \label{fig:pipeline-overview}
\end{figure}

We proposed a novel diversification method based on framing aspects, using the insights from the focus group. First, we describe the extraction pipeline, which supports the structure heuristic described in the results of the focus group session (Section \ref{sec:focus_group}). The pipeline forms the basis for the generation of recommendation lists that we use in the offline evaluation (Section \ref{sec:offline_evaluation}) and the online study (Section \ref{sec:online_study}). We implemented the pipeline using methods employed by the news aggregator platform and off-the-shelf natural language processing toolkits, such as IBM-Watson. We chose to use state-of-the-art and off-the-shelf methods used by the news aggregator platform to ensure output quality. Then we describe the distance function, which combines the metadata related to each framing aspect in a measure for viewpoint diversity for news articles. Finally, we present the re-ranking algorithm based on this viewpoint diversity measure. Our contribution, therefore, stands in the novelty of the overall diversification framework, rather than the implementation of specific components. Figure \ref{fig:pipeline-overview} shows an overview of the end-to-end pipeline.

\subsection{Metadata Extraction}
\label{sec:metadata_extraction}
For each framing aspect, as described in the definition of \citeauthor{entman1993framing}, we implemented an extraction pipeline:
        
        \paragraph{\Problem}
        As described in Section \ref{sec:related_work}, the problem definition can be understood as the central issue or topic under investigation \cite{matthes2008content}. Therefore, we decided to use a topic model as the main extraction method for this framing aspect. The model, provided by the research partner, included a 1000-topic latent Dirichlet allocation (LDA) model trained on 900k Dutch news articles. Based on the conclusions from the focus group described in Section \ref{sec:focus_group}, the title and the first \textit{x} paragraphs are used to retrieve metadata related to this framing aspect. We also applied multiple pre-processing steps on the content, including cleaning, chunking, tokenization, lemmatization and stop-word removal. 

        \paragraph{\Attribution + \Evaluation}
       According to \citet{entman1993framing}, the causal attribution of a frame relates to the forces creating the problem, while the moral judgements evaluate the causal attribution and their effect. From the discussion of the focus group session, described in Section \ref{sec:focus_group}, we concluded that the body of an article usually elaborates on these aspects. Additionally, paragraph-level seems to be the most suitable level of analysis. Therefore, a text-classification algorithm was applied using the IBM Watson Natural Language Processing API. The service returns a category for each paragraph according to a predefined five-level taxonomy, from the most general category (\emph{e.g.} level 1 - technology and computing), to the most specific one (\emph{e.g.}, level 5 - portable computer). To extract information related to the evaluation of these attributions, we also analyze the sentiment of these paragraphs, using the IBM Watson NLP API. Thereby, it would be able to identify if two articles evaluate the same aspects of a problem differently. The content of interest for this task includes all paragraphs except the $x$ introductory and $y$ concluding paragraphs. We optimize these variables during the offline evaluation.

         \paragraph{\Recommendation}
        Following the definition of \citet{entman1993framing}, a treatment recommendation suggests remedies for problems and predicts their likely effect. The research domain of suggestion mining, which involves the task of retrieving sentences that contain advice, tips, warnings and recommendations from opinionated texts \cite{negi2019suggestion}, was found to be highly relevant for this framing aspect \cite{negi2016study}. However, the state-of-the-art models are topic-specific \cite{negi2016study}, and can not be easily applicable to our domain. Thus, only the more naive rule-based approach could be applied for this study, being more generally applicable. In a crowdsourcing task with domain experts, we evaluated, and we optimized the generally applicable rules from the literature on the news article content. Afterwards, we implemented the method to extract sentences that contain suggestions from the article content. Then, to obtain comparable information between the suggestions of two articles, the suggestion sentences of each were classified using the same text-classification algorithm that was used for the causal attribution framing aspect. Corresponding to the conclusion of the focus group described in Section \ref{sec:focus_group}, the content of interest for this framing aspect includes the $y$ concluding paragraphs of an article. We optimize this variable in the offline evaluation.
        
\subsection{Distance Functions}
    \label{sec:distance_function}
    Having defined the extraction pipeline for each framing aspect, \emph{i.e.}, problem definition, causal attribution, moral evaluation and treatment recommendation \cite{entman1993framing}, we now define our distance function. We compare the extracted metadata for every pair of articles. Thus, we implement a distance function for each framing aspect.
      
\paragraph{\Problem}
        The metadata regarding the problem definition framing aspect involves a probability distribution over 1000 topics. Thus, we need a statistical distance measure. We chose the Kullback-Leibler divergence because it is one of the most commonly used statistical distance measures for LDA-models, and it is used in the comparable work \cite{tintarev2018same} on viewpoint diversification. 
        
        \paragraph{ \Attribution and \Evaluation}
       
        We compare the five-level taxonomy categories extracted from the pipeline described in the previous section, to obtain a distance measure for the causal attribution framing function of the primary frame. Thus, we use the weighted Jaccard index, which measures the similarity (or diversity) of two sets \cite{jaccard1901etude}. The index is calculated for each level of detail in the five-level taxonomy, such that we apply weight factors per taxonomy level. Thereby, overlap in higher levels of detail can contribute more to the overall similarity score. In the offline evaluation, we compare different weight factors per taxonomy-levels.
         
        For the moral evaluation framing aspect, we implement the distance function by multiplying the Jaccard distance and the absolute sentiment difference between each paragraph combination of two articles. Thus, paragraphs with no overlapping categories yield a value of zero, while highly similar paragraphs, with different sentiment scores, lead to high levels of diversity related to the moral evaluation framing aspect.
      
      \paragraph{\Recommendation} For the treatment recommendation we used the five-level taxonomy classification, \emph{i.e.}, from the most general to the most specific category, as returned by IBM Watson Natural Language Processing API, and the Jaccard index.

\subsection{Re-ranking}
    
We implement the re-ranking of the input list of articles using the Maximal Marginal Relevance (MMR) algorithm \cite{carbonell1998use}. In our case, the re-ranking consists of ranking news articles that are more diverse higher. First, we normalize the output of the distance functions related to each framing aspect using a min-max normalization, and then we combine them in a diversity score through a weighted sum. We optimize the weight factors during the offline evaluation. We note here that we re-rank news articles that are known to also be relevant for the given topic.

Where most re-ranking algorithms for recommender systems order lists only on relevance, the MMR algorithm provides a linear combination between diversity, in our case viewpoint diversity, and relevance, set by the parameter $\lambda$. Thus, the re-ranking algorithm is defined as follows:
     \begin{equation}
        MMR \equiv max_{i \in R \setminus S}[\lambda(Rel(i)-(1-\lambda)max_{j \in S}(1-Div(i||j))]
    \end{equation}
    
    Since this work proposes a measure for viewpoint diversity rather than a relevance measure, we decided to implement the relevance score using a simple frequency-inverse document frequency (TF-IDF) score.

\section{Offline Evaluation}
\label{sec:offline_evaluation}

In this section, we describe the offline evaluation of our viewpoint diversity-driven approach for re-ranking lists of news articles. 

\subsection{Materials}
For our offline experiment, we used the news dataset introduced in Section \ref{sec:dataset}, which covers 214 news articles on four topics.

\subsection{Procedure}
The experimental procedure consists of four main steps that we detail as follows. First, we process and enrich all the news articles in our dataset according to the four framing aspects as defined by \citet{entman1993framing}: problem definition, causal attribution, moral evaluation, and treatment recommendations (for details see Section \ref{sec:metadata_extraction}).

Second, we generate the diversity matrix by comparing all combinations of two articles, based on the enrichment described in Section \ref{sec:metadata_extraction}. Thus, using the distance function defined in Section \ref{sec:distance_function} we measure the dissimilarity of two articles based on the framing aspects. Finally, since the MMR algorithm re-ranks a list of news articles based on a linear combination between diversity and relevance, we calculate the TF-IDF relevance matrix, including a relevance score for each two article combination.

Third, we optimize the model variables and evaluate the performance using cross-validation. For each article $i$ in the dataset, we calculate a set of $s$ recommendations by re-ranking the remainder articles in the dataset. To prevent over-fitting, we use cross-validation. Thus, we split the dataset into $k$ distinct sets. We experimented with different values of $k={5,10,20}$ and $s={3,6,9}$. For every set, we take the following steps:
\begin{enumerate}[noitemsep,topsep=0pt]
\item  Grid search of model variables on training set: The training set contains the $k-1$ subsets of articles. We obtain the optimal combination of the model variables for the training set using a grid search. An overview of the model variables can be found in Table \ref{table:model-parameters} and in Section \ref{sec:model_variables}.
\item Evaluation on test set: After the variables are trained on the $k-1$ subsets, the model is evaluated on the test set for different values of $\lambda$, between 0 and 1 with a step of 0.1. As described before, for each article in the test set, a set of $s$ recommendations is calculated by re-ranking the remaining articles in the dataset. 
\end{enumerate}

And finally, we combined the results of all $k$ cross-validations.

\subsubsection{Model variables}
\label{sec:model_variables}
Table \ref{table:model-parameters} shows the model variables that we optimize during the offline evaluation. We choose the variation of the weights for each framing aspect such that no single framing aspect can have the majority. Additionally, a step-size of 0.1 is assumed to bring enough variation. We consider two variations for the taxonomy level weights: equal weights for each taxonomy level or ascending weights. Finally, the number of introductory and concluding paragraphs can be either $1$ or $2$.

\begin{table}[!ht]
    \centering
    \caption{Overview of possible values of model variables }
    \label{table:model-parameters}
    \resizebox{1\columnwidth}{!}{
        \begin{tabular}{l l} 
            \textbf{Variable} & \textbf{Values}\\
            \toprule
            Weight Framing function - Problem Definition & [0.1, 0.2, 0.3, 0,4]*\\
            \hline
            Weight Framing function - Causal Attribution & [0.1, 0.2, 0.3, 0,4]*\\
            \hline
            Weight Framing function - Moral Evaluation & [0.1, 0.2, 0.3, 0,4]*\\
            \hline
            Weight Framing function - Treatment Recommendation & [0,1, 0.2, 0.3, 0,4]*\\
            \hline
            Taxonomy level weight & [equal, ascending]\\
            \hline
            Number of introducing paragraphs & [1, 2]\\
            \hline
            Number of concluding paragraphs & [1, 2]\\
            \hline
            $\lambda$ & [0.0, 0.1, ..., 0.9]\\
            \bottomrule
        \end{tabular}
    }
        \begin{tablenotes}
          \small
          \item *Note that all framing function weight factors should sum up to 1
        \end{tablenotes}
\end{table}

\subsection{Evaluation Metrics}
We assess the performance of the viewpoint diversification method using a metric from literature \cite{tintarev2018same}. The metric is based on the Intra-List Diversity metric \cite{zhang2008avoiding,ziegler2005improving,tintarev2018same,vargas2011rank} and it is defined as the average distance between all pairs of articles $i$ and $j$, such that $i \neq j$. Thereby, the distance between a pair is defined by the articles' channels (predefined taxonomy of 20 high-level topics) and the articles' LDA topic-distribution, as derived from the enrichment methods in Section \ref{sec:methodology}:
\begin{equation}
    Distance(i,j) = 0.5 \times Distance_{Channels} +  0.5 \times Distance_{LDA}
\end{equation}

The channel distance is calculated using the cosine distance, whereas the LDA distance is computed using the Kullback-Leibler divergence.

\subsubsection{Additional metrics}
Besides the viewpoint diversity metric, we also measure the effectiveness of the diversification model on other properties, as follows:

\emph{Relevance}: We measure the TF-IDF relevance for the recommendation lists, such that we can measure the effectiveness of the viewpoint diversification method.

\emph{Kendall's $\tau$}: We compute the \textit{Kendall's $\tau$ rank correlation coefficient} \cite{kendall1948rank} to measure the similarity between two ranks of recommended items.

\emph{Average number of words}: We compute the average number of words for the recommended article lists as a measure of quality (\emph{i.e.}, longer news articles can be considered to be higher quality).

\emph{Publisher Ratio}: We measure the publisher ratio for the recommendation lists because this could potentially provide insights on the effect of the content diversity on the source diversity.

\subsection{Baseline}
To assess if the proposed diversification method can increase the viewpoint diversity based on the presented metric, we compare it with a baseline, consisting of a full relevance MMR, where $\lambda=1$, such that we rank the recommendations purely on the TF-IDF relevance. We chose this baseline because it has minimal effects on the recommendations in terms of viewpoint diversity.

    \begin{figure*}[!ht]
        \centering
        \begin{subfigure}[b]{0.24\textwidth}
            \centering
            \includegraphics[width=\textwidth]{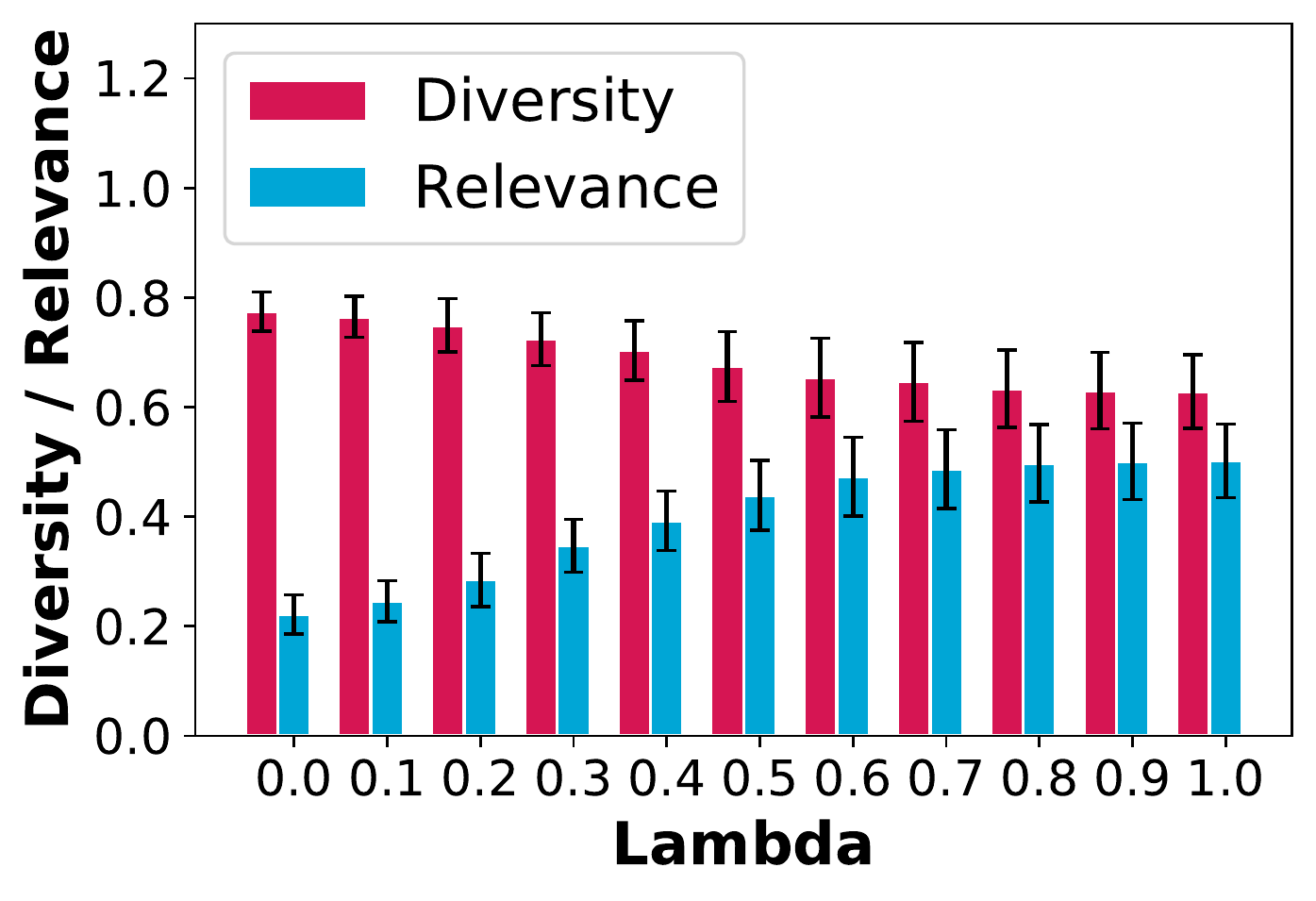}
            \caption[]%
            {{\small Topic: Black Lives Matter}}    
            \label{fig:offline-results-viewpoint-blm}
        \end{subfigure}
        \hfill
        \begin{subfigure}[b]{0.24\textwidth}  
            \centering 
            \includegraphics[width=\textwidth]{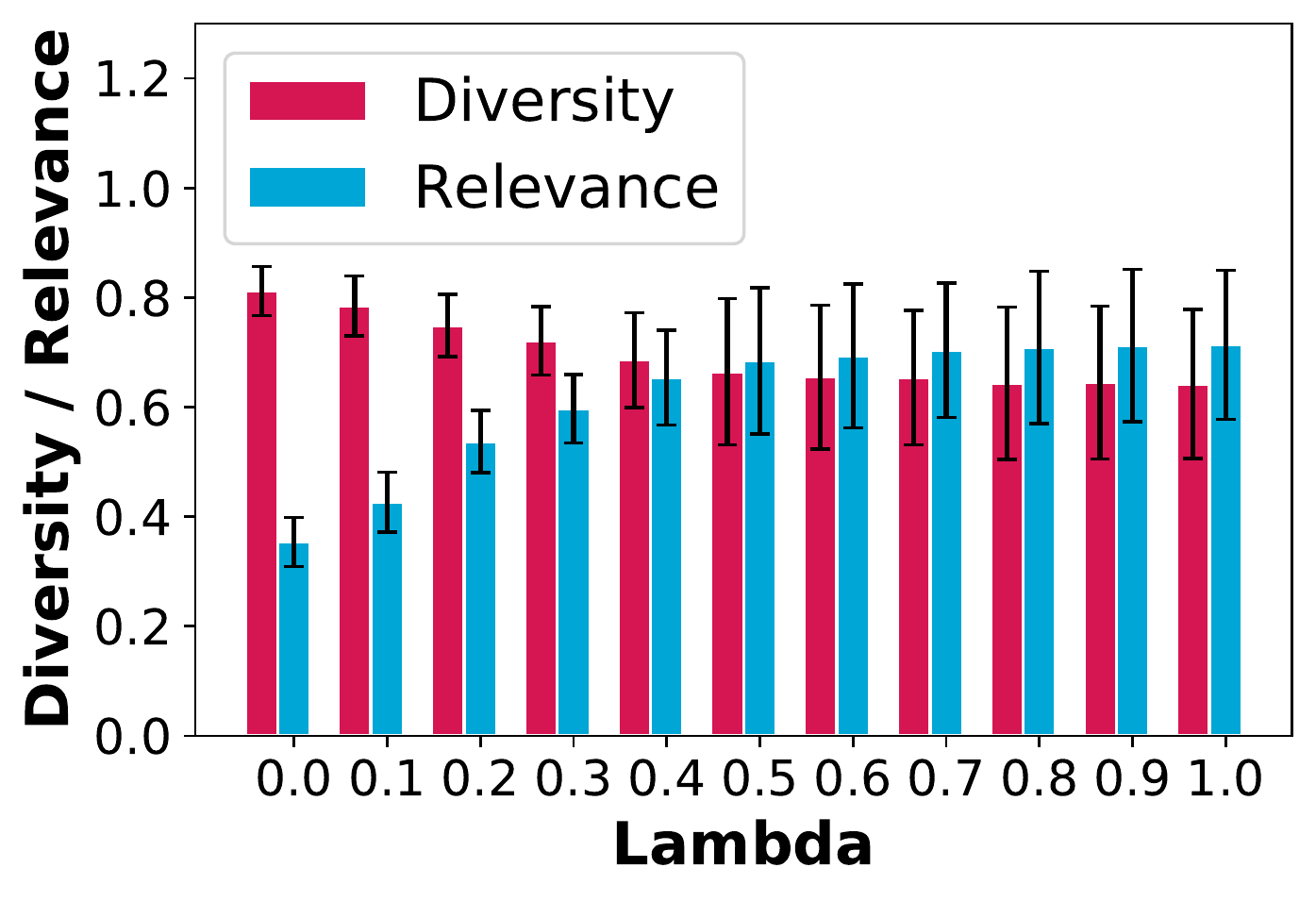}
            \caption[]%
            {{\small Topic: Coronavirus}}    
            \label{fig:offline-results-viewpoint-corona}
        \end{subfigure}
        \hfill
        \begin{subfigure}[b]{0.24\textwidth}   
            \centering 
            \includegraphics[width=\textwidth]{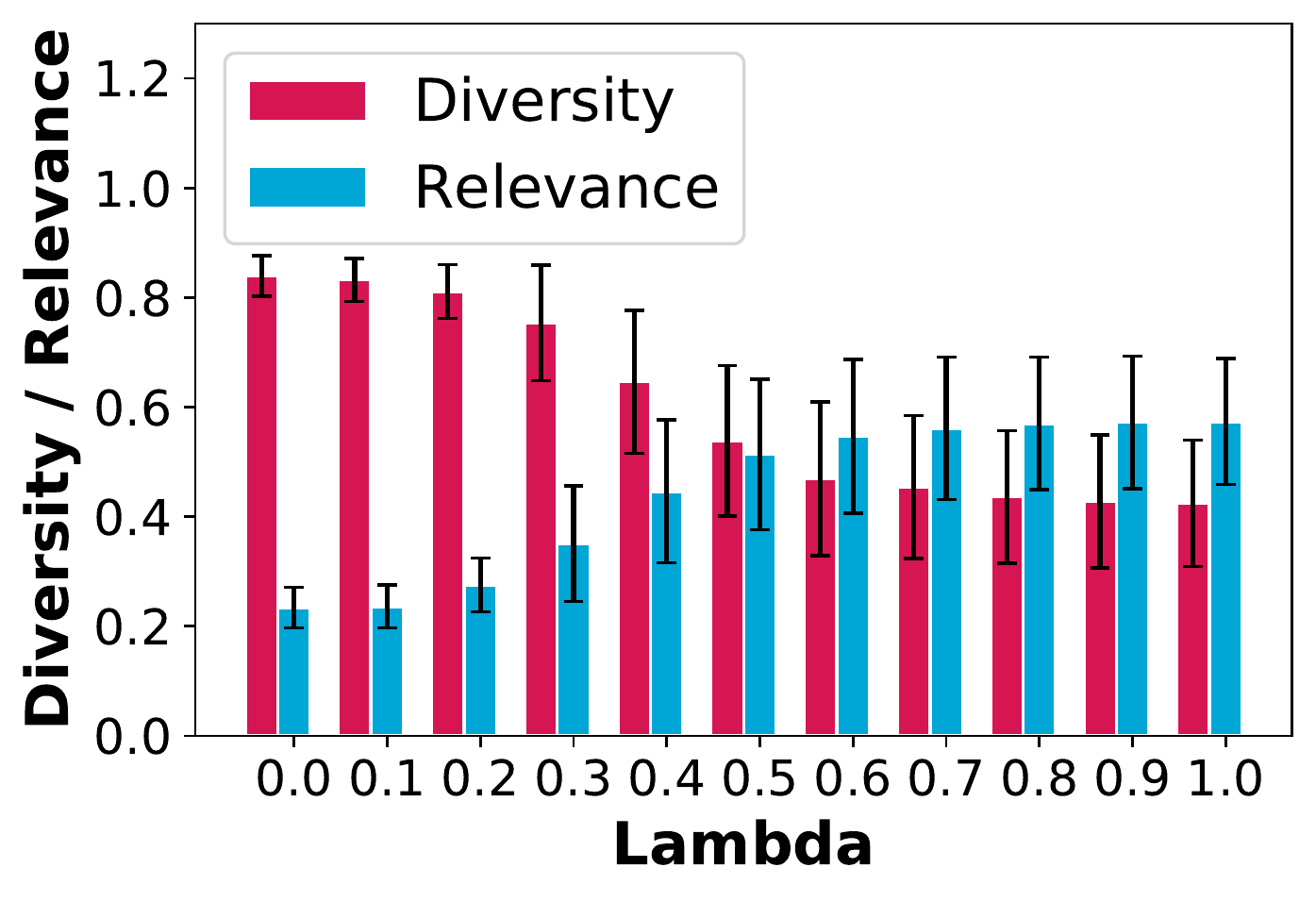}
            \caption[]%
            {{\small Topic: U.S. Elections}}    
            \label{fig:offline-results-viewpoint-verkiezingen_vs}
        \end{subfigure}
        \hfill
        \begin{subfigure}[b]{0.24\textwidth}   
            \centering 
            \includegraphics[width=\textwidth]{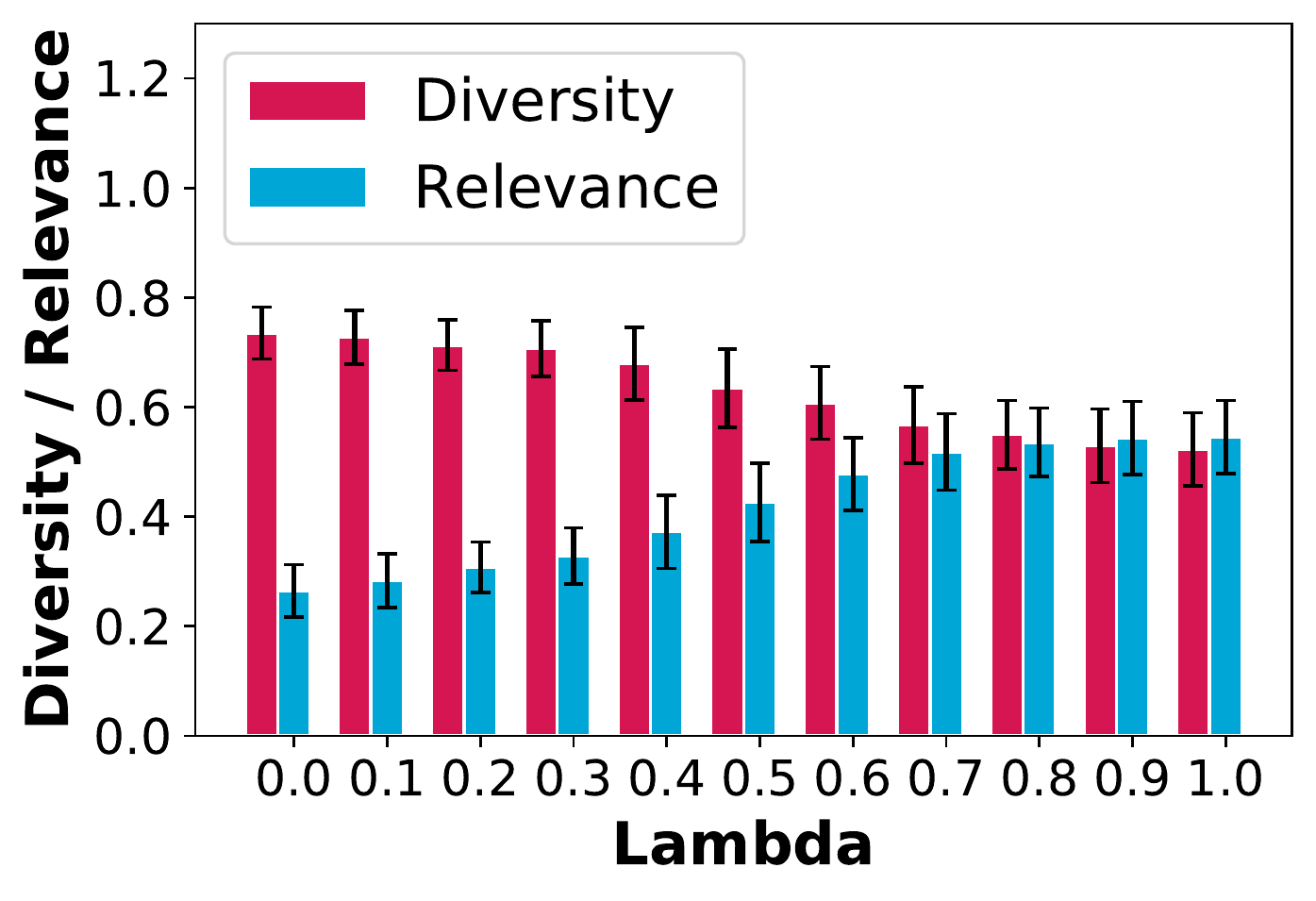}
            \caption[]%
            {{\small Topic: Big Tech}}    
            \label{fig:offline-results-viewpoint-bigtech}
        \end{subfigure}
        \caption[]
        {\small Diversity and relevance scores for different values of $\lambda$ per topic. } 
        \label{fig:offline-results-viewpoint}
    \end{figure*}
    
     \begin{figure*}[!ht]
      \centering
        \begin{subfigure}[c]{0.24\textwidth}
          \includegraphics[width=\textwidth]{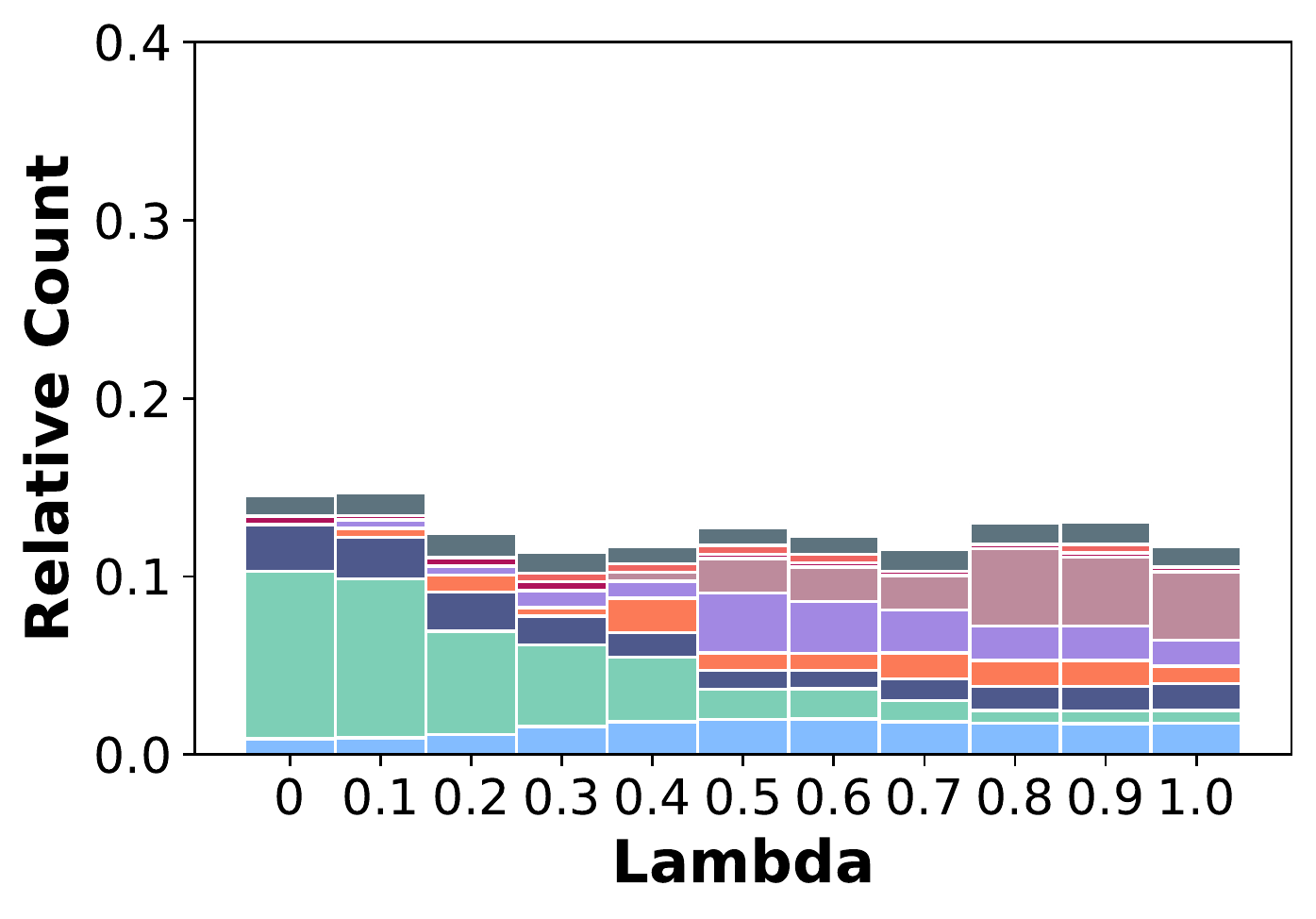}
          \caption{Topic: Black Lives Matter}
        \end{subfigure}
        \begin{subfigure}[c]{0.24\textwidth}
          \includegraphics[width=\textwidth]{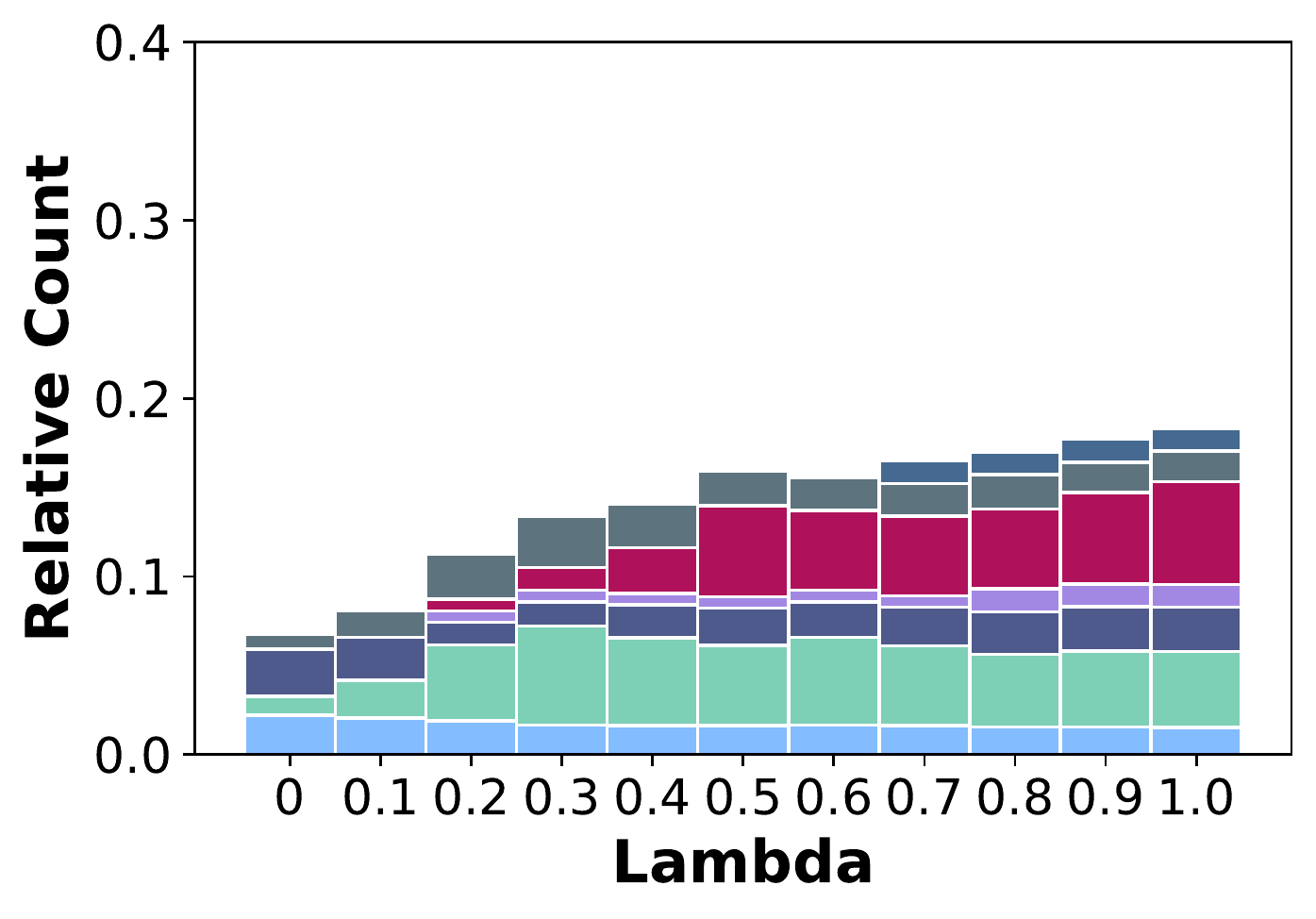}
          \caption{Topic: Coronavirus}
        \end{subfigure}
        \begin{subfigure}[c]{0.24\textwidth}
          \includegraphics[width=\textwidth]{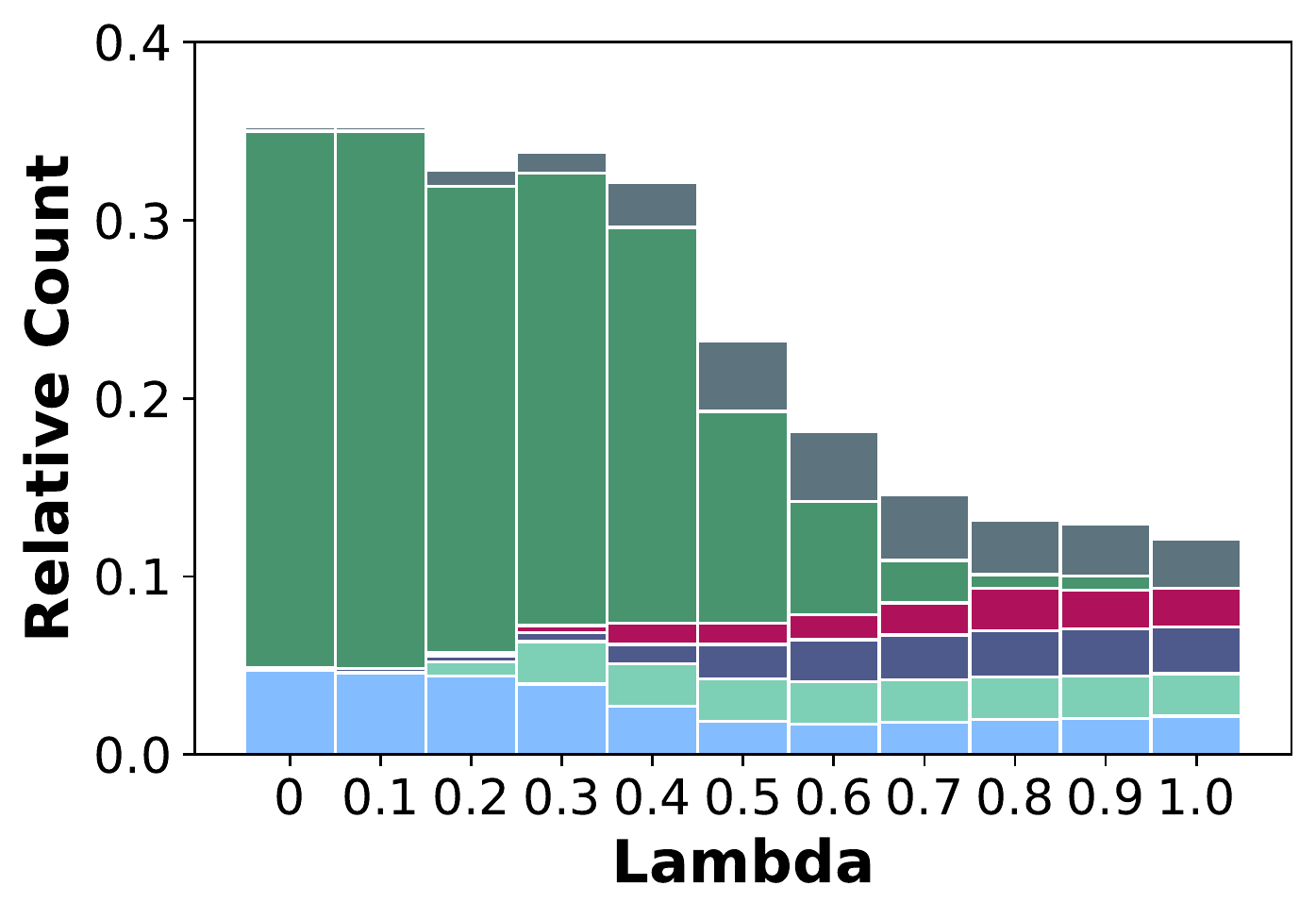}
          \caption{Topic: U.S. Elections}
        \end{subfigure}
        \begin{subfigure}[c]{0.24\textwidth}
          \includegraphics[width=\textwidth]{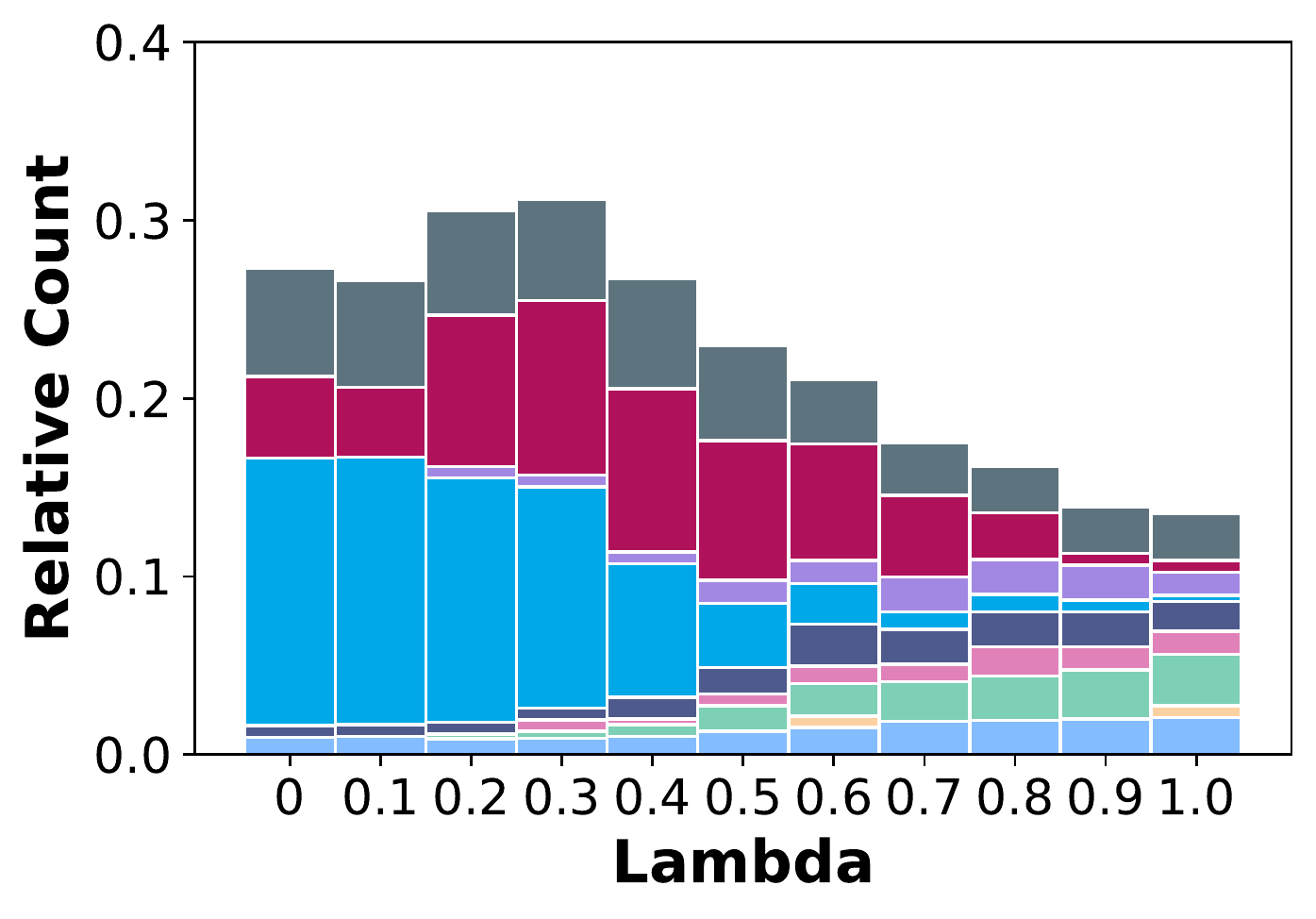}
          \caption{Topic: Big Tech}
        \end{subfigure}\\
      \caption{Average number of publishers in recommendation lists, normalised by the input ratio, for all topics.}
      \label{fig:offline-results-publishers}
    \end{figure*}

\subsection{Results}
In Figure \ref{fig:offline-results-viewpoint}, we show the performance of the model in terms of viewpoint diversity and relevance for different values of $\lambda$, and the optimal setting of the model variables. The red bars represent the results of the viewpoint diversity metric, while the blue bars represent the relevance scores. Variations of the cross-validation variable $k$ did not yield significant differences between the results, and thus, we fixed $k=10$. The list size $s$ did show to influence the number of publishers included in the recommended list, but the results were not significant. Thus, we fixed the list size to $s=3$, to better align the offline evaluation set up with the online evaluation set up, where only 3 recommended news articles can be shown at a time. Table \ref{tab:optimal_model_variables} shows the optimal model variables values, per topic.

Across all topics, the proposed diversification method is capable of increasing the viewpoint diversity of recommendation lists. According to the metric, the viewpoint diversity increases on average from 0.55 to 0.79 between $\lambda=1$ and $\lambda=0$. Additionally, the average relevance score decreases from 0.58 to 0.27.

\begin{table}[!ht]
        \centering
        \caption{Overview of model variables used during the offline and online evaluation for each topic: cross validation folds ($k$), recommended list size ($s$), number of introductory paragraphs, number of concluding paragraphs, general weights for the four framing aspects, category weights and $\lambda$. 
        }
        \resizebox{1\columnwidth}{!}{
        \label{tab:optimal_model_variables}
            \begin{tabular}{lccccccccc} 

                Topic & $k$ & $s$ & \begin{tabular}[c]{@{}c@{}}intro. \\ par.\end{tabular} & \begin{tabular}[c]{@{}c@{}}concl. \\par\end{tabular} & \begin{tabular}[c]{@{}c@{}}general \\weight\end{tabular} & \begin{tabular}[c]{@{}c@{}}cat. \\weight\end{tabular} & $\lambda$ \\
                \toprule
                \multicolumn{1}{c}{\begin{tabular}[l]{@{}l@{}}Black Lives Matter\end{tabular}} & 10 & 3 & 2 & 1 &[0.2, 0.4, 0.1, 0.3]& eq & 0 \\ \hline
                Coronavirus & 10 & 3 & 2 & 1 & [0.1, 0.4, 0.1, 0.4] & eq & 0\\
                \hline
                 U.S. Elections & 10 & 3 & 1 & 2 & [0.1, 0.4, 0.1, 0.4] & eq & 0\\
                \hline
                Big Tech  & 10 & 3 & 1 & 2 & [0.2, 0.4, 0.1, 0.3] & asc & 0\\
                \bottomrule
            \end{tabular}
            }
    \end{table}

\paragraph{Kendall's $\tau$} We computed the Kendall's $\tau$ rank correlation to assess whether the proposed diversification method is capable of providing different recommendation lists compared to the baseline. We computed the coefficient between the baseline ($\lambda=1$) and each other value of $\lambda=[0.0,0.1,...,0.9]$. Overall, we observed that the re-ranking of the set of recommendations based on viewpoint diversity results in different recommendation lists compared to the baseline. The coefficient decreases for smaller values of $\lambda$, but it is bounded around $\tau=0$ for decreasing values of $\lambda$. 

\paragraph{Average number of words} We observe no consistent pattern in the average number of words for different values of $\lambda$ across topics. For the \emph{Black Lives Matter} and \emph{Big Tech} topics, the average number of words increases for larger values of $\lambda$, for the \emph{U.S. Elections} topic the average decreases and for \emph{Coronavirus} the average is stable.

\paragraph{Publisher ratio} Figure \ref{fig:offline-results-publishers} shows the average number of articles in the recommended lists, normalized by the input ratio, for each value of $\lambda$. For every topic, the number of publishers increases for larger values of $\lambda$ and the number of different publishers for the baseline recommendation list is larger than the one in the diverse recommendation list. Thus, we observe that the diversification method influences the publisher ratio. For small values of $\lambda$, some publishers get amplified, while others are excluded. We see this effect primarily for the topics of \emph{U.S. Elections} and \emph{Big Tech}. The topic of \emph{Corona Virus} seems to be the only exception.

\section{Online Study}
\label{sec:online_study}

We conducted a between-subjects online study on the Blendle platform to compare the reading behaviour of users who receive news articles optimized only for relevance, versus news articles that are also diverse on viewpoint.

\subsection{Materials}
\label{sec:materials_online}

In the online study, we used the articles collected in Section \ref{sec:dataset}.

\subsection{Participants}
\label{sec:participants}

We selected 2076 active users of the news aggregator platform. 
These users were assumed to most likely see and use the recommendation functionality. 
We included only users who clicked at least four times on a recommended article below any article read, in the last 14 days before the study. Groups for baseline and diversified recommendations were created by randomly splitting the users.

\subsection{Independent Variables}
\label{sec:independent_variables_online}

In the between-subjects user study we manipulated the following conditions, referring to the recommended list of news articles:
\begin{itemize}[noitemsep,topsep=0pt]
    \item\textbf{ baseline recommendation}: was implemented using a MMR that was based only on relevance ($\lambda=1.0$)
    \item \textbf{diversified recommendation}: was implemented using a MMR that maximized viewpoint diversity ($\lambda=0.0$)
\end{itemize}

\subsection{Procedure}
\label{sec:procedure}
During the two-week experiment, six days per week, we provided recommendations for two articles featured on the selected users' homepage. We provided sets of three recommendations below the content on the reading page of the original article. Every morning, we chose these two articles manually, to match any of the topics that we selected (\emph{Black Lives Matter}, \emph{Big Tech}, \emph{Coronavirus}, and \emph{U.S. Elections}). Afterwards, both the baseline and diversified recommendation sets were calculated for both articles and included in the news aggregator platform. 

\subsection{Dependent Variables}
\label{sec:dependent_variables_online}
To analyze the reading behaviour of the two different user groups and answer \textbf{RQ1}, we measure specific events on the news aggregator platform (\emph{i.e.}, check whether the user opened the article and if the user finished reading the article). Based on these available events, we observe multiple implicit (click-through-rate per news article, click-through-rate per recommendation set and completion rate of recommendation) and explicit (heart ratio) measures of the reading behaviour. To answer \textbf{RQ2}, we look into presentation characteristics of the recommended articles (\emph{i.e.}, presence of editorial title, presence of thumbnail and counting number of hearts).

\emph{1. Click-through rate per article:} The number of clicks on a news article is divided by the total number of users who finished one of the original news articles for which that article was recommended. The completion of an original news article is registered using a scroll-position.
        
\emph{2. Click-through rate per recommendation set:} The total number of clicks on either of the three news articles in the recommendation set is divided by the number of users who finished the original news article (using scroll-position) for which the recommendation set was presented. 
        
\emph{3. Completion rate of recommendation:} Is implemented as the number of users that read the full recommended article (using scroll-position) divided by the number of users who opened the news article. The completion rate is assumed to be a measure for the user satisfaction with the recommendations. We can argue that short news articles are more likely to be completed than long news articles. Thus, we also analyze the completion rate of a news article in relation to the number of words in the news article.


\emph{4. Favourite ratio:} The news aggregator platform allows users to mark an article as a favourite, illustrated by an icon of a heart. The users can click this icon at the end of the article content. We implemented the measure as the number of users of the user group (baseline or diverse) that clicked on the icon, divided by the number of users in the same group that completed the article. The metric is assumed to be a marker of user satisfaction with the article. 

\emph{5. Presentation characteristics:} We measured three additional properties of a recommended article during the experiment, which referred to the presentation characteristics of recommended news articles. First, the editorial team can replace the original title of a news article with a custom, editorial title. In general, these custom titles are longer and more explanatory than the original ones. Second, articles can be presented with or without a thumbnail image. Third, the number of users who selected the article as a favourite is visualised by a counting number of hearts in the left-upper corner of an article banner. All three properties are assumed to potentially influence the click-through rate and are, therefore, measured during the experiment. 

\emph{6. Source diversity:} Finally, we also measured the influence of the source diversity of the recommendation set on the click-through rate. As seen in Section \ref{sec:offline_evaluation}, higher levels of viewpoint diversity showed to influence the number of times a publisher is included in the recommendation.

\subsection{Results}
\label{sec:online-results}
    
The online study ran six days a week for two weeks. Thus, we provided recommendations below 24 articles. During the experiment, the topic of \emph{Coronavirus} became extremely prominent, so we provided recommendations below 18 out of 24 news articles on this topic. In contrast, the \emph{Black Lives Matter} topic lost all actuality, resulting in no recommendations for this topic. For the \emph{U.S. Elections} topic, we provided recommendations below four articles, and for the \emph{Big Tech} topic, below two news articles. 

\paragraph{Click-through rate per recommended article} The mean click -through rate per recommended article for the baseline recommendations was 0.11 (stderr. = 0.011) while for the diversified recommendations was 0.087 (stderr. = 0.0083) when looking at all topics. Furthermore, according to the Mann-Whitney U test (U=570, p-val$>$0.05), we did not find a significant difference between the two user groups in terms of click-through rate per recommended article. The same result holds per topic.

\begin{figure*}[!ht]
        \centering
        \begin{subfigure}[b]{0.23\textwidth}
            \centering
            \includegraphics[width=\textwidth]{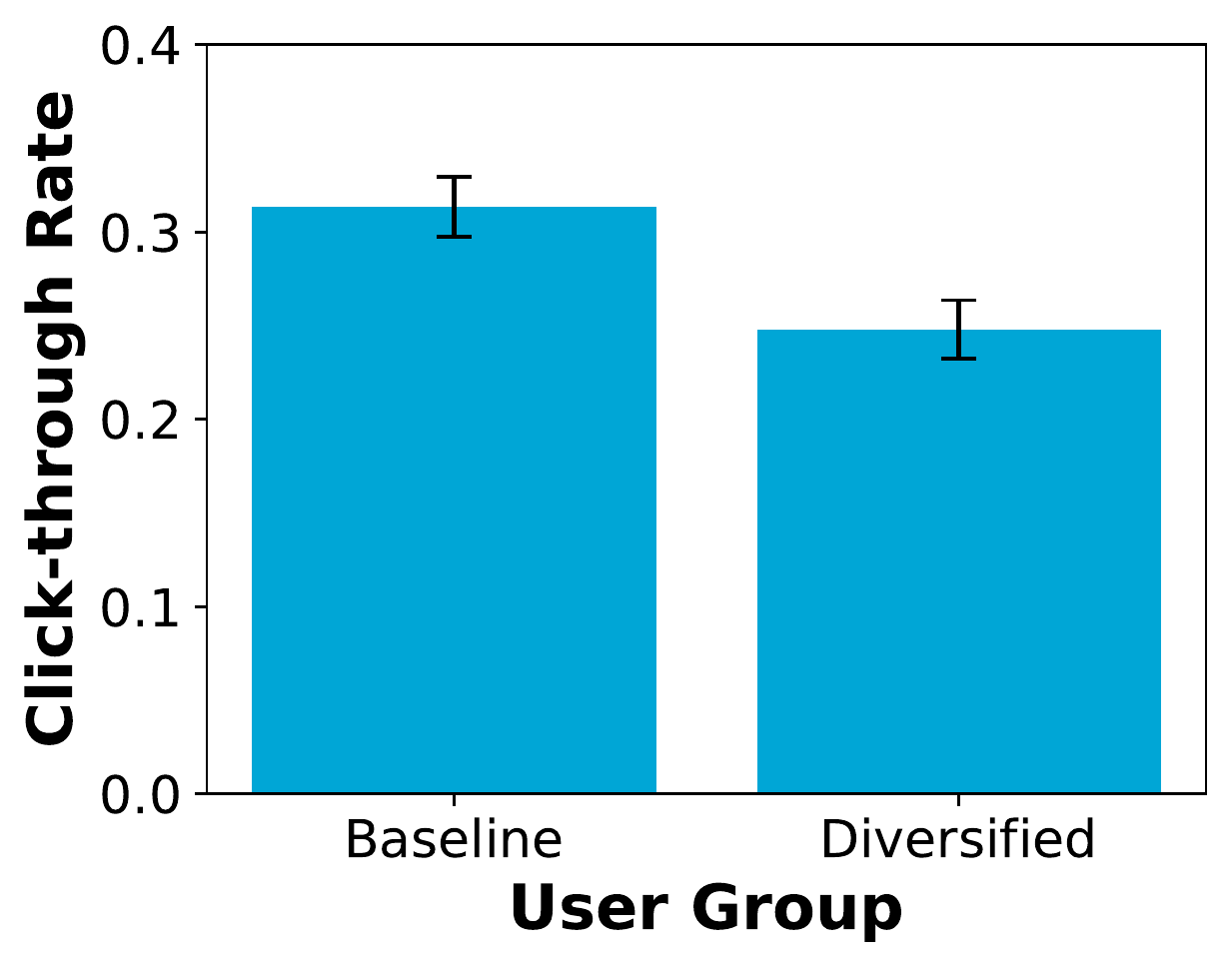}
            \caption{Click-through rate per recommended set, for the two user groups.}   
            \label{fig:online-results-clickthrough-total}
        \end{subfigure}
        \hfill
        \begin{subfigure}[b]{0.23\textwidth}  
            \centering 
            \includegraphics[width=\textwidth]{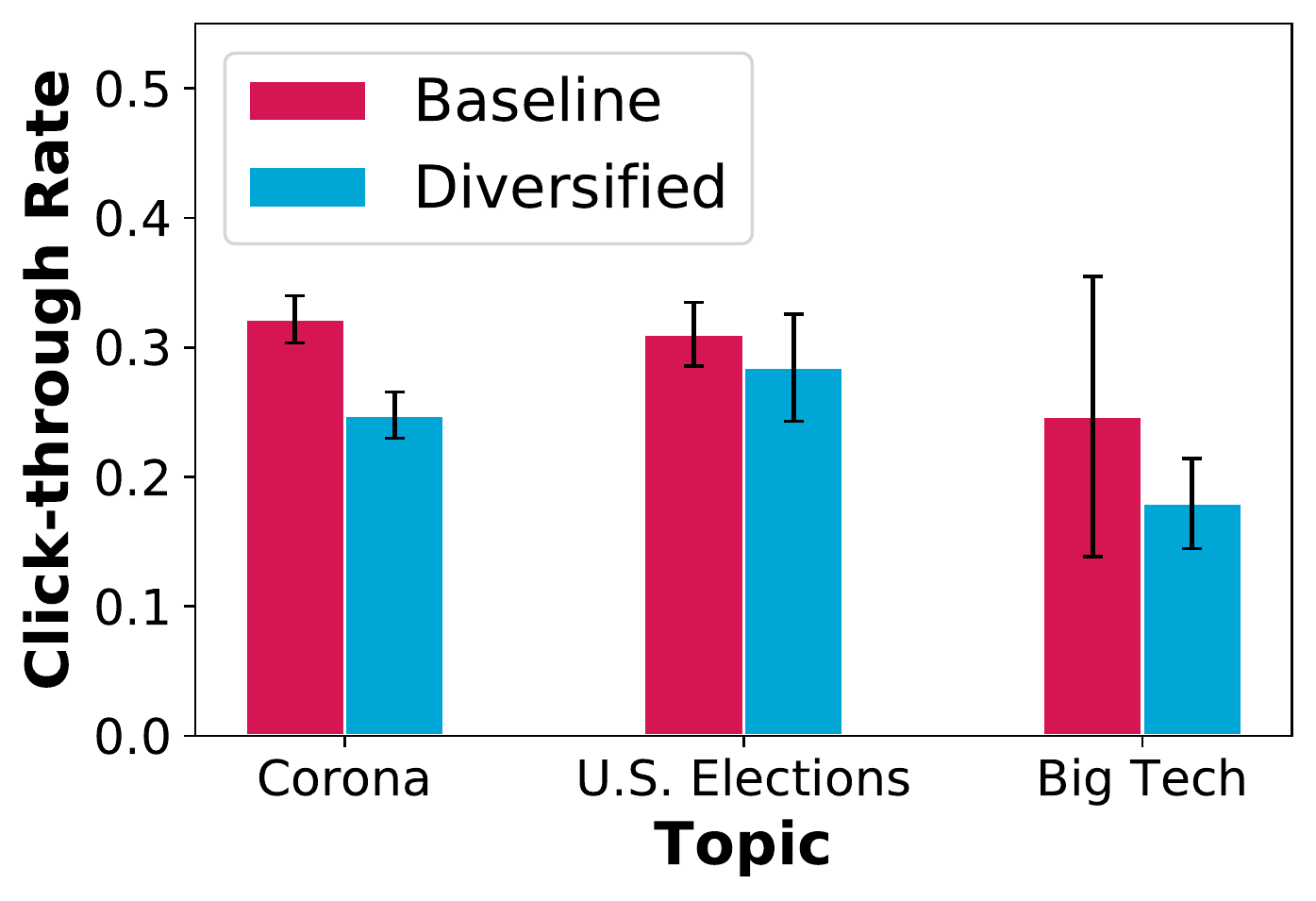}
            \caption{Click-through rate per recommended set and per topic, for the diversified user group.}    
            \label{fig:online-results-clickthrough-topic}
        \end{subfigure}
        \hfill
        \begin{subfigure}[b]{0.23\textwidth}   
            \centering 
            \includegraphics[width=\textwidth]{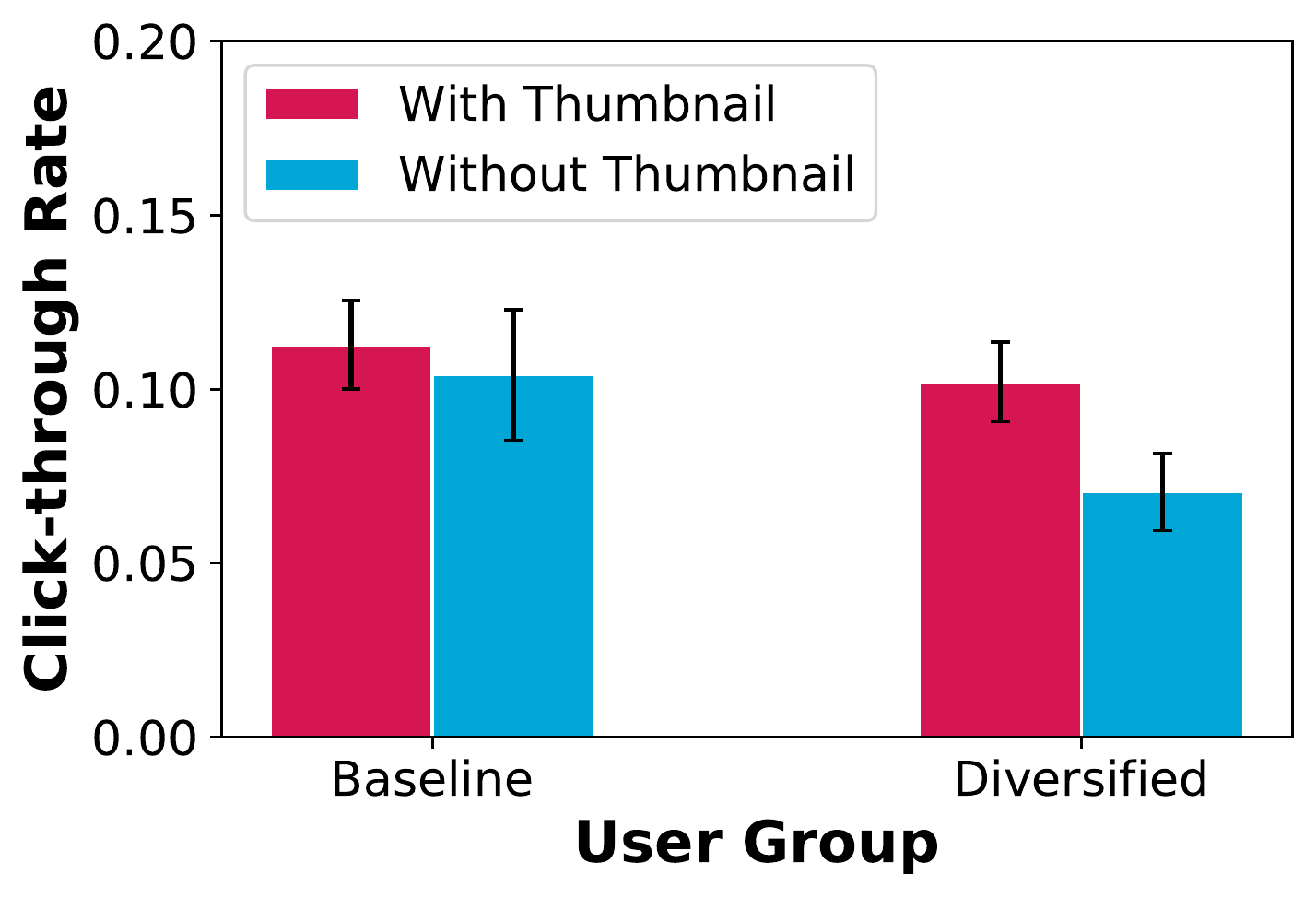}
            \caption{Influence of the thumbnail image as presentation characteristic, for the two user groups.}   
            \label{fig:online-results-thumbnail}
        \end{subfigure}
        \hfill
        \begin{subfigure}[b]{0.23\textwidth}   
            \centering 
            \includegraphics[width=\textwidth]{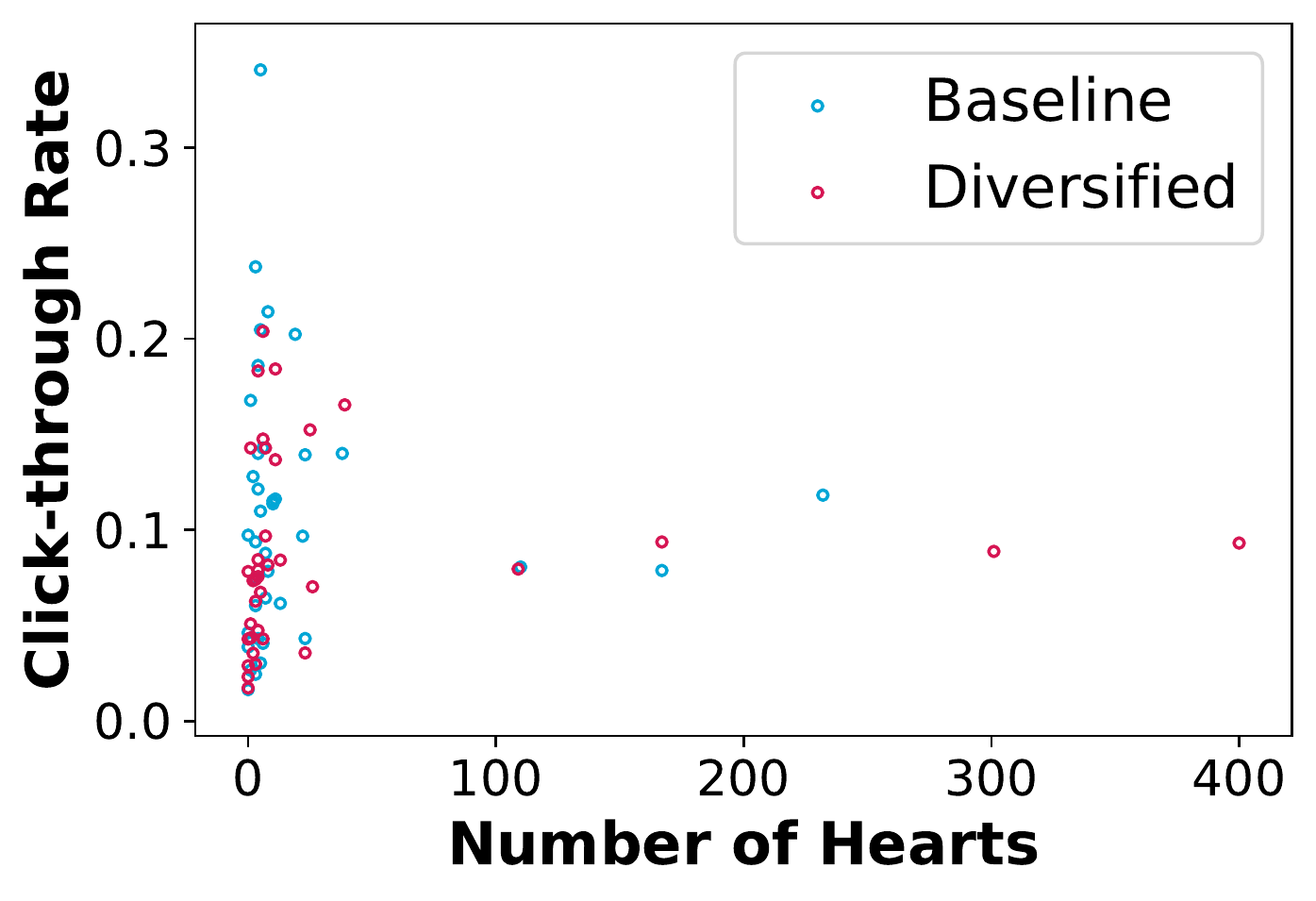}
            \caption{Influence of the hearts as presentation characteristic, for the two user groups.}
            \label{fig:offline-results-hearts}
        \end{subfigure}
        \caption{Overview of significant results in the online study}
        \label{fig:online-results}
    \end{figure*}

\paragraph{Click-through rate per recommended set} The mean click-through rate per recommended set for the baseline recommendations was 0.31 (stderr. = 0.016) while for the diversified recommendations was 0.25 (stderr. = 0.016) when looking at all topics (Figure \ref{fig:online-results-clickthrough-total}). According to the Mann-Whitney U test (U=2.9, p-val$<$0.05), we find a significant difference between the mean click-through rate per recommended sets for the two user groups. Per topic, we find such difference significant only for \emph{Coronavirus}, shown in Figure \ref{fig:online-results-clickthrough-topic}, with a click-through rate per recommended set of 0.32 (stderr. = 0.018) for the baseline recommendations and 0.25 (stderr. = 0.018) for the diversified recommendations (U=80.0, p-val$<$0.05). For the other topics, we found no significant difference between the two user groups. 

\paragraph{Completion rate} We found no significant difference in terms of completion rate for the two user groups. We also applied the Spearman's rank correlation to see whether the completion rate is correlated with the length of the articles. However, we did not find any correlation in either of the two conditions. 

\paragraph{Heart ratio} We found no significant difference, for all topics and across topics, in terms of heart ratio for the two user groups. This suggests that the quality of the recommendations was comparable between the two conditions. 



\subsubsection{Influence of presentation characteristics}

We measured the influence of three factors, namely the presence of an editorial title, the presence of a thumbnail and the number of users that chose the article as a favourite on the click-through rate of an article. 

\paragraph{Editorial title} Regarding the influence of the inclusion of an editorial title on the click-through rate, no statistical significance was found for neither user groups.

\paragraph{Thumbnail image} We found no significant influence of the inclusion of a thumbnail image on the click-through rate for baseline users. In contrast, recommendations
with a thumbnail are 3.1\% more times opened than recommendations without a thumbnail for diverse users, as seen in Figure \ref{fig:online-results-thumbnail}, a difference that is also statistically significant.

\paragraph{Favorite articles} We applied the Spearman's rank correlation to see whether we find a correlation between the click-through rate and the number of hearts. Figure \ref{fig:offline-results-hearts} shows the distribution of click-through rates and the number of hearts. We only found a moderate positive correlation of 0.57, also statistically significant (p-val$<<$0.05) for the diversified user group. 

\subsubsection{Source diversity}
As seen in the offline evaluation, higher levels of viewpoint diversity turned out to have remarkable effects on the publisher ratio. Therefore, we also evaluated the effect of the source diversity of a recommendation set on the click-through rate. For each recommendation set, we computed the number of different publishers and we found recommendation sets in which all articles are from a different publisher and sets in which two articles are from the same publisher. Afterwards, the click-through was calculated for each category. The results for both the baseline users and diverse users show that no statistically significant difference can be found in the click-through rate between two or three different publishers in the recommendation set for neither baseline nor diversified users.




\section{Discussion}
\label{sec:discussion}

We first discuss the results of the offline and online evaluation and then provide an overview of the limitations of our approach.

\subsection{Offline Evaluation}

The offline evaluation indicated that the proposed method is capable of increasing the viewpoint diversity of recommendation sets according to the metric defined in \cite{tintarev2018same}. The average viewpoint diversity scores across all topics increased from 0.55 to 0.79 for an increasing level of diversity in the MMR algorithm. Simultaneously, the average relevance score decreased from 0.58 to 0.27. Remarkably, the diversity score of 0.41 in \cite{tintarev2018same} is considerably smaller than the maximum average value of 0.79 found in this work. A possible factor could be the fact that in \cite{tintarev2018same} the LDA topic model was excluded from the diversification method to prevent any interference with the evaluation metric, whereas the diversification method in this work still depends on an LDA topic-model. Therefore, the difference in viewpoint diversity scores between the methods can possibly appear due to the interference of metadata between the viewpoint diversity metric and diversification method in this work.


A remarkable effect of the diversification algorithm that was found in the offline evaluation includes the decreasing publisher ratio for larger contributions of diversity in the MMR. After investigating the effect in more detail, it was found that the maximum frequency an article is included in the recommendation lists is around 2 to 4 times higher at $\lambda$ = 0, compared to $\lambda$ = 1. Thus, for larger contributions of diversity, the algorithm increasingly selects the same article for the recommendation lists. This could be a possible explanation of the decreasing publisher ratio, suggesting that some outliers in the dataset get amplified, thereby suppressing the inclusion of different sources. To be able to study this effect thoroughly, the offline evaluation could have benefited from a setup in which it was possible to assess the contribution of individual framing aspects to the global viewpoint diversity score per article. 


We can conduct a broader discussion about the viewpoint diversity metric used. 
Although approaches that use source diversity are more popular, scholars generally agree that viewpoint diversity can only be achieved by fostering content diversity, because, multiple sources can still refer to the same point of view \cite{voakes1996diversity}. Based on these findings, this study used a content-based approach. From the results of the offline evaluation, it became clear that increasing levels of content diversity exclude multiple publishers and thus, decreases source diversity. Moreover, some specific publishers got amplified remarkably for high levels of content diversity. Therefore, viewpoint diversification methods could benefit from considering both content and source diversity.

\subsection{Online Evaluation}
\label{sec:discussion-online}
No major influence of viewpoint diversification on the reading behaviour was found, except for the click-through rate calculated per recommendation set, which indicated a statistically significant difference between baseline and diverse users of 6.5\% (in favour for baseline recommendations). However, the results of the click-through rate calculated per recommendation indicated no significant difference between the two user groups. Likewise, the other two measurements of the reading behaviour, including the completion rate of recommendations and the ratio of users who selected a recommendation as a favourite, showed no significant difference between baseline and diverse users.

In reflection on the motivation of this study, the proposed diversification for news media is capable of enhancing the viewpoint diversity of news recommendation, while maintaining comparable measures of the reading behaviour of users. The results thus suggest that recommender systems are capable of preserving the quality standards of multiperspectivity in online news environments. Thereby, situations of extreme low diversity, known as filter bubbles, could also be mitigated.

These results are in contrast with the most comparable study, \citet{tintarev2018same}, who found a negative effect on intent to read diversified news articles. The authors proposed a viewpoint diversification method based on the MMR-algorithm with linguistic features, such as gravity, complexity and emotional tone. During a user study, 15 participants were asked to make a forced choice between a recommendation from the diverse set and a recommendation from the baseline set, after reading an article on the same topic. It was found that 66\% of the participants chose the baseline article, compared with 33\% who chose the diverse article. However, in the current study, we observed the reading behaviour of both user groups without them being aware, and we argue that the present setup simulates the situation in a more realistic way.

Additionally, the results shed light on the importance of how a recommendation is presented. Multiple presentation properties, such as the inclusion of a thumbnail image and the number of times an article is marked as favourite, were shown to have a significant influence on the click-through rate of recommendations. Future research, thus, should not only address the capability of a model to enhance viewpoint diversity according to an offline metric but also evaluate what presentation characteristics could impact the users' willingness to read multiperspectival news. Related research on \textit{viewpoint-aware interfaces}, which aim to explain the recommendation choices to users, can be seen as very valuable \cite{tintarev2017presenting,nagulendra2014understanding}.

\subsection{Limitations}
\label{sec:limitations}
We further discuss the limitations of our approach. 


\textit{Choice of participants in the online study.} Only users who frequently followed recommendations below articles were selected for the experiment. Thus, the click-through rates presented in this study are higher than for average news readers. 


\textit{Limited number of topics and articles.} 
For both the online and offline evaluations, we used only opinion pieces. Furthermore, each evaluation had a limited number of topics, namely four, as well as a limited number of news articles. New topics could reveal additional results that hold across topics. 



\textit{Missing user perceptions.} While we were able to study user behavior at a reasonable scale, a notable omission is users' qualitative judgement of viewpoint diversity in the resulting recommendations. We plan to continue collaborating with the news aggregator platform to refine the proposed framework, \emph{i.e.}, to improve the viewpoints extraction.

\textit{Presentation characteristics.} 
Some presentation characteristics, and in particular the heart ratio, could also be markers of quality. Further qualitative analysis is needed to \emph{e.g.}, understand how much of user behavior is directed by quality. We also saw that for some topics the presence of thumbnail was more common than for other topics, and it would be relevant to study whether this also interacted with user perceptions of relevance or quality.


\textit{Relevance metric.} The offline study could use a more sophisticated relevance measure between the recommendation and the original article. The relevance score was based on a simple TF-IDF score, limited to the terms in a handcrafted search query.


\textit{Influence of $\lambda$.} 
Given limited time for online testing, we only compared against 
a maximum viewpoint diversity score.

\textit{Influence of publishers.}
In Figure \ref{fig:offline-results-publishers} we see that, although 15 publishers are represented in the datasets, three publishers are predominant. Due to the limited number of articles and the unbalance in terms of publishers, the inclusion of a wide variety of perspectives on a topic can be challenged. 

\section{Conclusions}
\label{sec:conclusions}


In this paper, we proposed a novel method for enhancing the diversity of \textit{viewpoints} in lists of news recommendations. Inspired by research in communication science, we identified frames as the most suitable conceptualization for news content diversity. 
We operationalized this concept as a computational measure, and we applied it in a re-ranking of topic relevant recommended lists, to form the basis of a novel viewpoint diversification method. 

In an \textit{offline evaluation}, we 
found that the proposed method improved the diversity of the recommended items considerably, according to a viewpoint diversity metric from literature. We also conducted an \textit{online study} with more than 2000 users, on the Blendle platform, a Dutch news aggregator. 
 The reading behaviour of users receiving diversified recommendations was largely comparable to those in the baseline. Besides, the results suggest that presentation characteristics (thumbnail image, and the number of hearts) lead to significant differences in reading behaviour. 
These results suggest that research on presentation aspects for recommendations may be just as relevant as novel viewpoint diversification methods, to achieve multiperspectivity in automated online news environments.

As future work, we plan to investigate further the presentation characteristics and how they influence user experience, in addition to behaviour. In more controlled settings, we will study the relative effects of actual (e.g., as judged by experts) versus perceived quality (e.g., number of hearts in the interface) of recommended news items. 
Future work will also focus on defining a better metric to measure viewpoint diversity, as opposed to topic diversity, c.f.,  \cite{drawsassessing}. 
Additionally, 
we learnt that contextual information, \emph{i.e.}, general knowledge about a topic (\emph{e.g.}, the current measures in place to stop the spread of coronavirus) can also be essential to reveal a specific frame. 
We hope that this work will encourage further research on how framing can be defined, conceptualized, and evaluated in the computational domain.


\bibliographystyle{ACM-Reference-Format}
\balance
\bibliography{facct_main}


\end{document}